\documentclass[conference]{IEEEtran}
\IEEEoverridecommandlockouts
\usepackage{cite}
\usepackage{amsmath,amssymb,amsfonts}
\usepackage{graphicx}
\usepackage{amsmath}
\usepackage{textcomp}
\usepackage{booktabs} 
\usepackage{siunitx} 
\usepackage{subcaption}
\usepackage{caption}
\usepackage{soul}
\usepackage{array} 
\usepackage[linesnumbered,ruled,vlined]{algorithm2e}

\usepackage{algpseudocode}
\usepackage[numbers]{natbib}
\usepackage{xcolor}
\def\BibTeX{{\rm B\kern-.05em{\sc i\kern-.025em b}\kern-.08em
    T\kern-.1667em\lower.7ex\hbox{E}\kern-.125emX}}

\usepackage{mdframed}
\newmdenv[
  backgroundcolor=gray!10, 
  linewidth=0.5pt, 
  innertopmargin=6pt, 
  innerbottommargin=6pt, 
  skipabove=6pt, 
  skipbelow=6pt 
]{graybox}
\begin{document}
\author{\IEEEauthorblockN{Hongyu Ke}
\IEEEauthorblockA{\textit{Department of Computer Science} \\
\textit{Georgia State University}\\
Atlanta, USA \\
hke3@student.gsu.edu}
\and
\IEEEauthorblockN{Wanxin Jin}
\IEEEauthorblockA{\textit{Ira A. Fulton Schools of Engineering} \\
\textit{Arizona State University}\\
Tempe, USA \\
wanxin.jin@asu.edu}
\and
\IEEEauthorblockN{Haoxin Wang}
\IEEEauthorblockA{\textit{Department of Computer Science} \\
\textit{Georgia State University}\\
haoxinwang@gsu.edu}
}
\title{CarbonCP: Carbon-Aware DNN Partitioning with Conformal Prediction for Sustainable Edge Intelligence}

\maketitle

\begin{abstract}
This paper presents a solution to address carbon emission mitigation for end-to-end edge computing systems, including the computing at battery-powered edge devices and servers, as well as the communications between them. We design and implement, \textsc{CarbonCP}, a context-adaptive, carbon-aware, and uncertainty-aware AI inference framework built upon conformal prediction theory, which balances operational carbon emissions, end-to-end latency, and battery consumption of edge devices through DNN partitioning under varying system processing contexts and carbon intensity. Our experimental results demonstrate that \textsc{CarbonCP} is effective in substantially reducing operational carbon emissions, up to 58.8\%, while maintaining key user-centric performance metrics with only 9.9\% error rate.
\end{abstract}


\section{Introduction}
\label{sc:intro}

Recently, the implementation of artificial intelligence (AI) and machine learning (ML) at the network edge has emerged as a promising solution for bringing computing and storage resources closer to users, thereby meeting the stringent latency requirements of AI applications.
Today, edge computing has progressively permeated a variety of mainstream service domains, including smart cities, healthcare, manufacturing, agriculture, and transportation. Considerable research efforts have been undertaken to accelerate the latency and enhance the accuracy of AI inference on the edge.
\textit{However, environmental sustainability, particularly in terms of the carbon emissions from the overall edge computing system (including the computing at battery-powered devices and edge servers as well as the communications between them), is under-explored, yet it is becoming increasingly significant as the deployment of these systems expands.}

\textbf{Environmental Sustainability in Edge Intelligence: Why It Matters.} Carbon abatement is critically important in addressing the escalating threat of climate change. As of 2019, the energy consumed by information and computing technologies (ICT) accounts for $2\%$ of global carbon emissions, half that of the aviation industry \cite{freitag2021real}. This sustainability challenge is further exacerbated by the widespread adoption of AI-empowered edge devices, the super-linear growth in the complexity of AI models, and the exponential increase in demand for computing and networking capabilities. For instance, the volume of data used for AI has increased by $2.4\times$, resulting in a $3.2\times$ increase in the demand for data ingestion bandwidth \cite{wu2022sustainable}. Moreover, the size of AI models used for language translation tasks has expanded by $1000 \times$ \cite{hernandez2020measuring}, leading to a substantial increase in the demand for computing resources at the edge. Additionally, emerging edge applications, such as augmented reality that requires a consistent latency of less than $33$ ms, have led to significant increases in computational and communication power in edge computing systems, often at the expense of their operational carbon footprint.
Therefore, it is imperative that our research aligns with emission reduction targets and focuses on identifying and developing innovative, sustainable solutions to address the significant sustainability challenges in edge intelligence.

\begin{figure}[t]
\centerline{\includegraphics[width=0.42\textwidth]{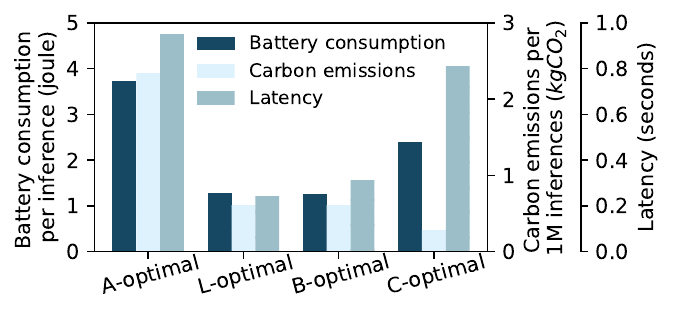}}
\caption{Trade-offs among the operational carbon emissions of the end-to-end edge computing system, end-to-end latency, and battery consumption. Lower is better.}
\vspace{-0.25in}
\label{fig:intro_adp}
\end{figure}

\begin{figure*}[t]
  \centering
  \begin{subfigure}[b]{0.49\textwidth}
    \centering
    \includegraphics[width=\linewidth]{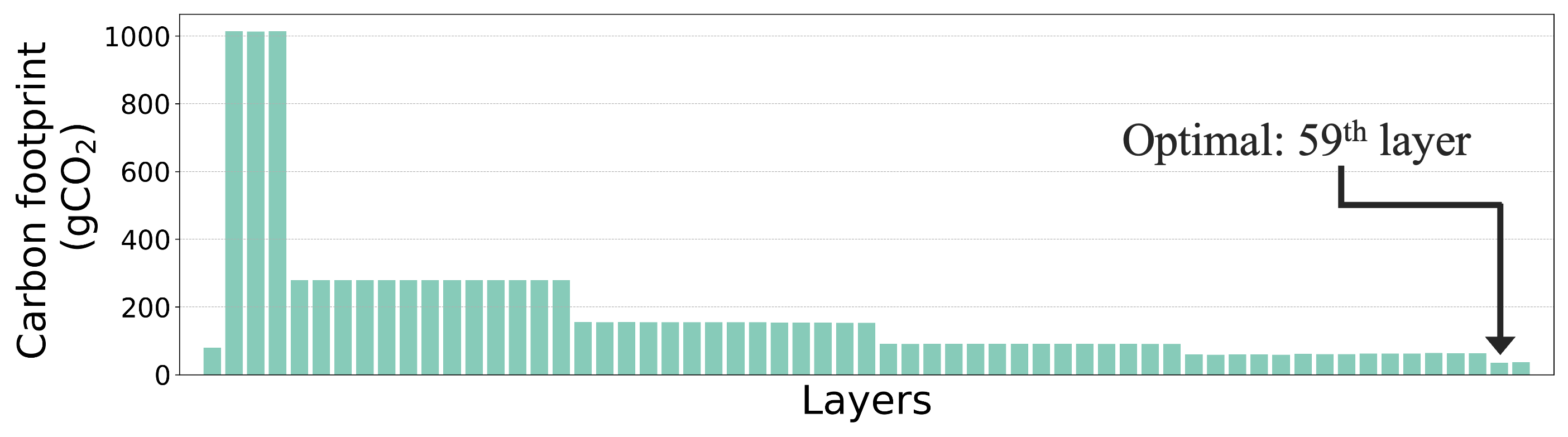}
    \caption{Carbon-minimal DNN partitioning under processing context 1}
    \label{fig:sub1}
  \end{subfigure}
  \hfill
  \begin{subfigure}[b]{0.49\textwidth}
    \centering
    \includegraphics[width=\linewidth]{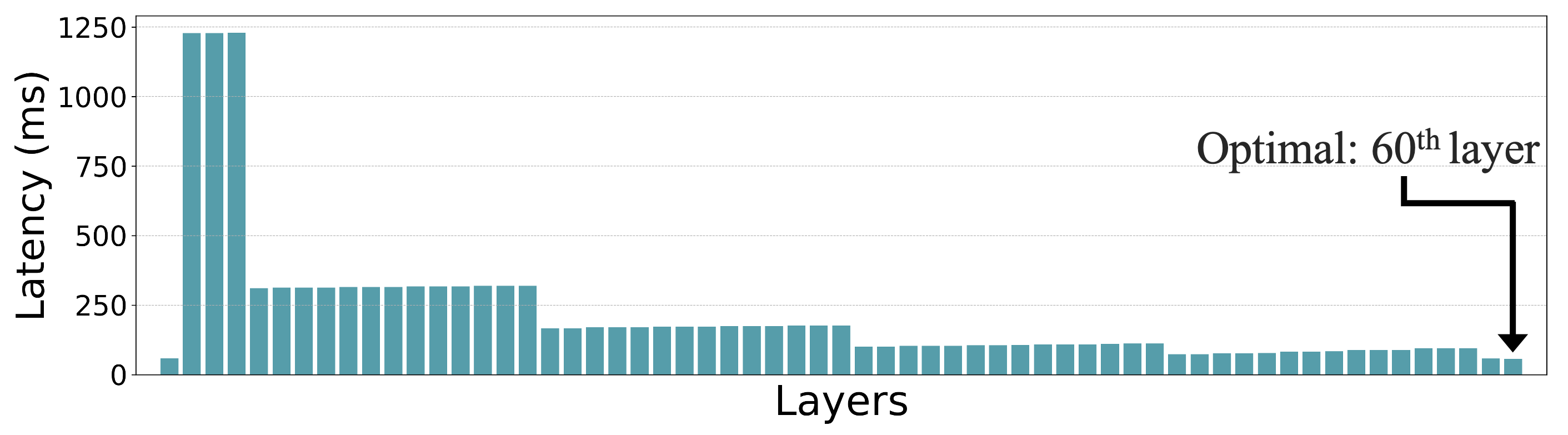}
    \caption{Latency-minimal DNN partitioning under processing context 1}
    \label{fig:sub2}
  \end{subfigure}
  \hfill
  \begin{subfigure}[b]{0.49\textwidth}
    \centering
    \includegraphics[width=\linewidth]{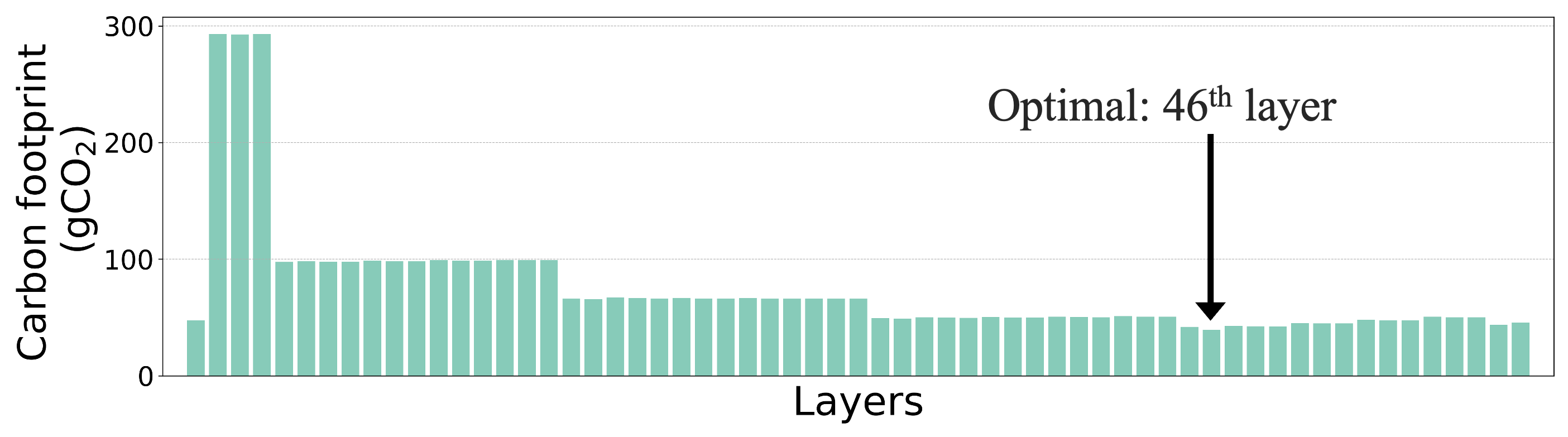}
    \caption{Carbon-minimal DNN partitioning under processing context 2}
    \label{fig:sub3}
  \end{subfigure}
    \begin{subfigure}[b]{0.49\textwidth}
    \centering 
    \includegraphics[width=\linewidth]{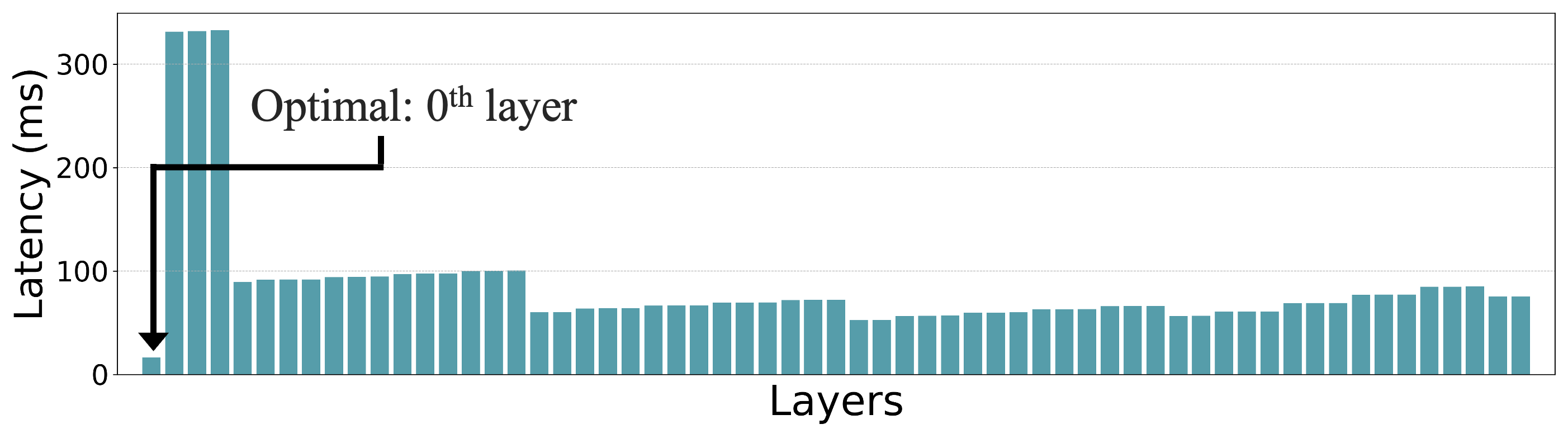}
    \caption{Latency-minimal DNN partitioning under processing context 2}
    \label{fig:sub4}
  \end{subfigure}
  \caption{Impact of dynamic processing contexts on DNN partitioning solutions, achieving minimal overall operational carbon emissions vs. minimal end-to-end latency in edge computing systems.}
  \label{fig:contextVariation}
\end{figure*}

To this end, \textit{the goal of this paper is to design a novel carbon-aware AI inference framework that incorporates deep neural network (DNN) partitioning to mitigate the overall operational carbon emissions of an end-to-end edge computing system} (including the computing at battery-powered devices and edge servers, as well as the communications between them). Unfortunately, the integration of carbon-awareness into AI inference processes often compromises other critical user-centric properties, such as latency, inference accuracy, and energy consumption in battery-powered edge devices. To demonstrate these trade-offs, we implement four system and network configurations driven by distinct objectives: maximizing inference accuracy (A-optimal), minimizing end-to-end latency (L-optimal), minimizing edge device battery consumption (B-optimal), and minimizing overall operational carbon emissions (C-optimal).
Performance results are presented in Figure \ref{fig:intro_adp}. C-optimal achieves the lowest overall operational carbon emissions but dramatically increases battery consumption and end-to-end latency by $90.5\%$ and $161.3\%$, respectively, compared to B-optimal. This suggests that optimizing for carbon emissions from the end-to-end edge computing system generally leads to significantly higher battery consumption and increased latency, as it requires more computations on low-power edge devices. 

Additionally, these complex trade-offs can be further exacerbated by the dynamics of the processing context within the edge computing system and environment. Such dynamics might include variations in carbon intensity, computational resources on edge devices (e.g., GPU and CPU utilization and frequencies), available network bandwidth between edge devices and the server, and server computational resources (e.g., GPU utilization). We have conducted experimental studies to demonstrate these impacts, which will be discussed in Section~\ref{sc:preliminary}.
\textit{Consequently, it is imperative yet challenging to effectively minimize overall operational carbon emissions from the edge computing system without compromising critical user-centric performance metrics.}


\textbf{Our Contributions Towards Sustainable Edge Intelligence.} In this paper, we study these research challenges and design a novel AI inference framework for edge computing systems, named \textsc{CarbonCP} that dynamically balances overall operational carbon emissions, end-to-end latency, and battery consumption of edge devices through adaptively optimizing the DNN partition solutions in response to varying system processing contexts and carbon intensity. Our proposed \textsc{CarbonCP} framework adopts a practical and effective uncertainty estimation method built upon Conformal Prediction (CP) theory \cite{shafer2008tutorial}, which obtains an interval for the DNN partition point that is stochastically guaranteed to contain the accurate optimal solution, adhering to a confidence level (\S\ref{ssc:confidence}) preferred by the user. 

In summary, our main contributions are as follows:
\begin{itemize}
    \item To the best of our knowledge, this study is the first to systematically explore the design of environmentally sustainable edge computing systems, presenting experimental evidence that highlights the opportunities and trade-offs in DNN partitioning for carbon abatement.
    \item We design and implement \textsc{CarbonCP}, a \textit{context-adaptive, carbon-aware, and uncertainty-aware} AI inference framework based on CP, specifically aimed at minimizing overall operational carbon emissions while maintaining key user-centric performance metrics, such as latency and battery consumption of edge devices.
    \item We extensively evaluate our approach, and the experimental results demonstrate that \textsc{CarbonCP} can significantly reduce carbon emissions by accurately predicting the optimal DNN partition solutions.
\end{itemize}

\section{Preliminary Experiments and Key Insights}
\label{sc:preliminary}

In this section, we describe our preliminary experiments designed to evaluate the impact of various factors on operational carbon emissions and user-centric performance metrics within an edge computing system. These experimental results yield three key insights that motivate the design of our \textsc{CarbonCP} inference framework for edge intelligence.

\subsection{The Impact of Dynamic Processing Context on DNN Partitioning}

Figure \ref{fig:contextVariation} visualizes the impact of processing context dynamics on determining the optimal DNN partitioning within the edge computing system, aimed at minimizing overall operational carbon emissions or end-to-end latency. In these experiments, we employ ResNet-18 on edge devices and a GPU server.
Figures \ref{fig:sub1} and \ref{fig:sub3} compare the optimal DNN partition solutions with the least overall operational carbon emissions under two distinct processing contexts. We observe that the most carbon-efficient partition point significantly shift depending on the processing context, as shown by the transition from the 59th layer to the 46th layer. Figures \ref{fig:sub3} and \ref{fig:sub4} demonstrate the optimal DNN partition solutions that cater to different objectives under consistent . The optimal solutions for minimizing latency are different from those for minimizing carbon emissions, indicating a trade-off between latency and environmental sustainability. For instance, Figure \ref{fig:sub4} shows that the 0th layer is preferred for the lowest latency, whereas Figure \ref{fig:sub3} reveals that this does not offer the lowest operational carbon emissions.

While previous work has leveraged DNN partitioning primarily for improving latency and inference accuracy \cite{hu2019dynamic, li2019edge, eshratifar2019jointdnn}, to the best of our knowledge, it has not yet been exploited for carbon reduction within the context of edge computing. 
The complex variations in optimal partition solutions, as depicted in these results, suggest the importance of a sophisticated approach to balance the competing demands of latency and carbon emission reduction.
This balance becomes especially challenging in \textsc{CarbonCP}'s case because the system framework must dynamically respond to the varying processing context\footnote{In this paper, the system processing context includes: computational resources on edge devices (e.g., GPU and CPU utilization and frequencies), available network bandwidth between edge devices and server, and server computational resources (e.g., GPU utilization)} and carbon intensity of the energy source (\S\ref{ssc:ci}).

\begin{graybox}
\noindent\textbf{Insight 1.} Partitioning DNNs offers opportunities to reduce the overall operational carbon emissions with edge computing systems by enabling more efficient use of computational and communication resources. However, the inherent trade-offs between latency and carbon emissions, along with the dynamic processing contexts, pose challenges in exploiting this opportunity, thereby necessitating the integration of \textbf{contextual adaptation} into the design of the \textsc{CarbonCP}.
\end{graybox}

\begin{figure}[t]
\centerline{\includegraphics[width=0.48\textwidth]{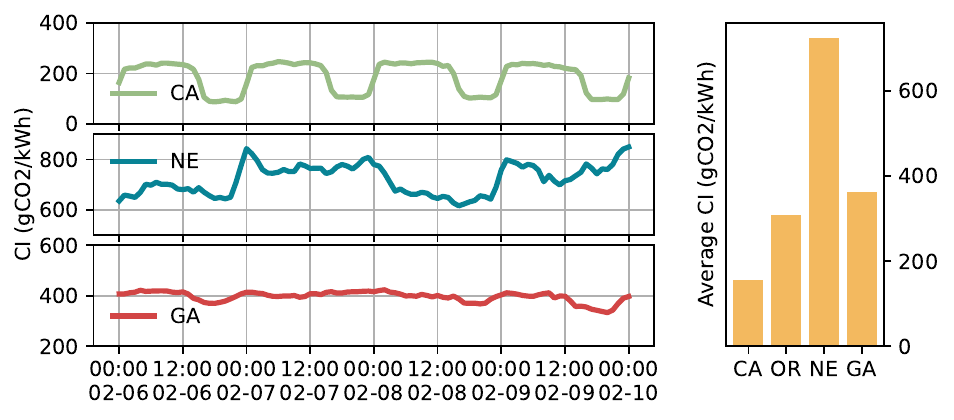}}
\caption{Carbon intensity (CI) varies both spatially and temporally in the United States \cite{Emaps}.}
\vspace{-0.15in}
\label{fig:task1ci}
\end{figure}

\subsection{The Impact of Carbon Intensity on DNN Partitioning}
\label{ssc:ci}
The carbon intensity is a critical metric that measures the ``greenness'' of the generated energy, quantifying the amount of carbon dioxide (CO\textsubscript{2}) emissions produced per unit of energy consumed (e.g., gCO\textsubscript{2}/kWh). In the United States, carbon intensity varies both spatially and temporally.
We evaluate the carbon intensity of all states in the United States and visualize the data of four representative states in Figure \ref{fig:task1ci}. We observe that there is considerable variation among different states in terms of the characteristics of carbon intensity, specifically regarding their average values, variance, and patterns of variation. For instance, California's (CA) carbon intensity exhibits a clear diurnal pattern, fluctuating within a 24-hour period. Typically, carbon intensity peaks during the night and early morning due to the decreased solar generation and rising demand.
However, Nebraska's (NE) carbon intensity exhibits more abrupt changes and lacks the predictable pattern observed in CA. Particularly, while fluctuations in carbon intensity occur in NE, they do not appear to be systematically linked to the time of day. This could be attributed to the fact that the primary renewable energy source in NE is wind power, which, unlike solar energy, can provide a more consistent output irrespective of the time of day or weather conditions.
Additionally, Georgia's (GA) carbon intensity remains relatively stable, largely due to the very low contribution of renewable energy, which stands at only 7.6\%.

\begin{figure}[t]
  \centering
  \begin{subfigure}[b]{0.48\textwidth}
    \centering
    \includegraphics[width=\linewidth]{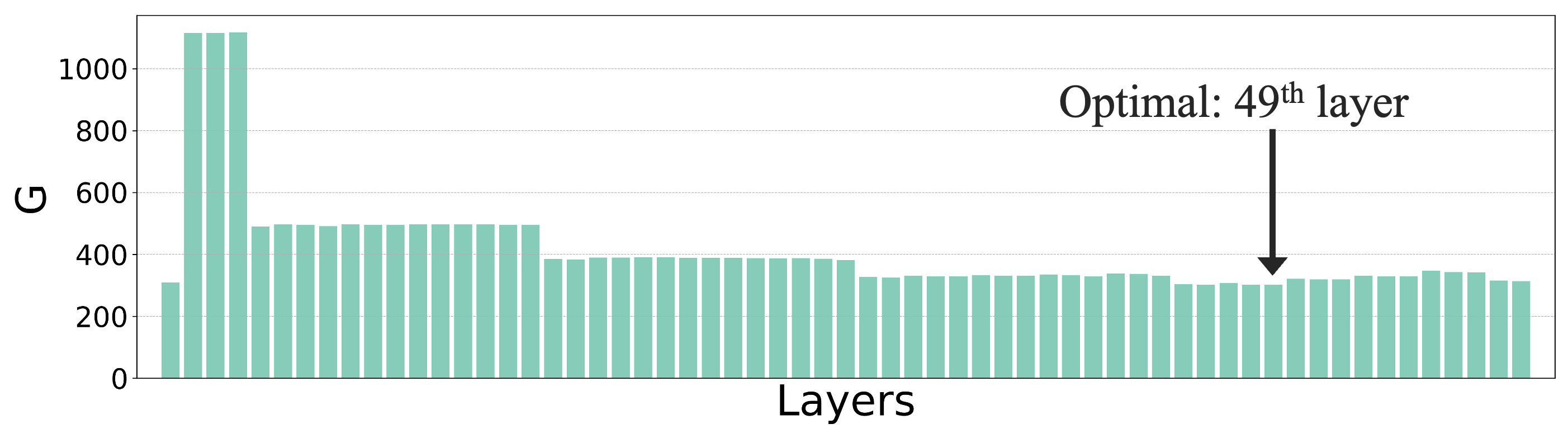}
    \caption{$G$-optimized DNN partitioning with $CI = 600$ gCO\textsubscript{2}/kWh}
    \label{fig:sub201}
  \end{subfigure}
  \hfill
  \begin{subfigure}[b]{0.48\textwidth}
    \centering
    \includegraphics[width=\linewidth]{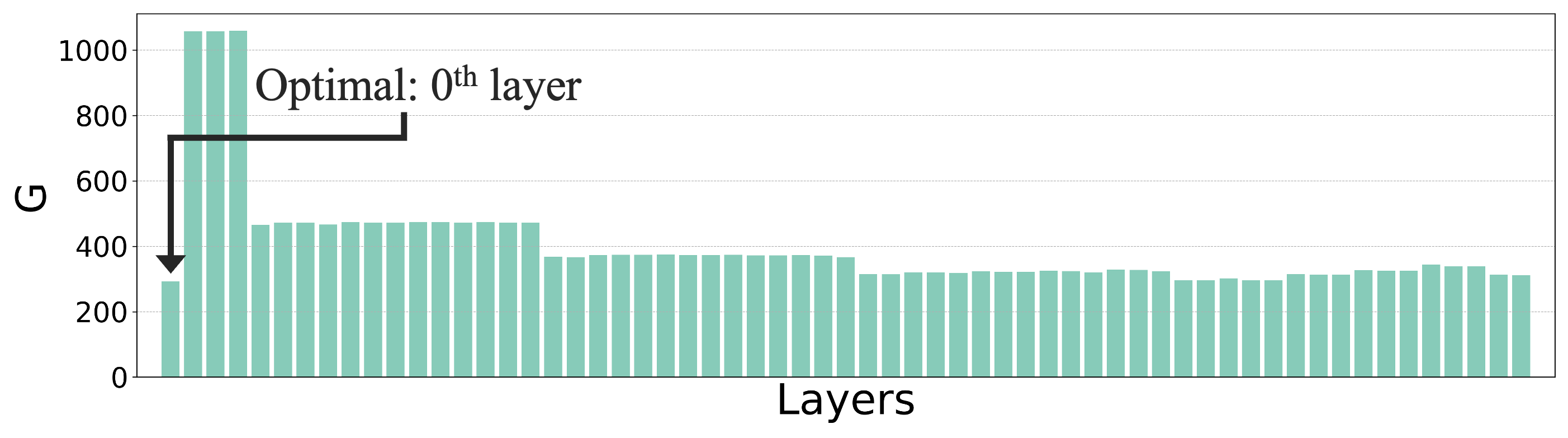}
    \caption{$G$-optimized DNN partitioning with $CI = 549$ gCO\textsubscript{2}/kWh}
    \label{fig:sub202}
  \end{subfigure}
  \caption{Impact of carbon intensity on DNN partitioning.}
  \vspace{-0.15in}
  \label{fig:CIVariation}
\end{figure}

These observations suggest that the carbon intensity is a function of geographical location where the edge computing system is deployed and operated, and the time during its operation. Variations in carbon intensity can introduce new design space for optimizing overall operational carbon emissions through strategic DNN partitioning.
To support our idea, we evaluate the optimal DNN partitions under two different carbon intensity scenarios. Furthermore, we define a new objective, denoted as $G$, aimed to strike a balance the battery consumption (a critical user-centric metric) of edge devices and overall operational carbon emissions. The results, depicted in Figure \ref{fig:CIVariation}, demonstrate that the optimal partition point shifts from the 49th to the 0th layer when carbon intensity is reduced from $600$ to $549$ gCO\textsubscript{2}/kWh.

\begin{graybox}
\noindent\textbf{Insight 2.} Carbon intensity exhibits both spatial and temporal variability. The design of the \textsc{CarbonCP} framework must, therefore, exploit this opportunity to adapt the DNN partitioning both geographically and over time, incorporating \textbf{carbon awareness} to improve environmental sustainability.  
\end{graybox}

\subsection{Lack of Uncertainty Assessment in DNN Partitioning}



Most existing studies on DNN partitioning typically employ a pre-trained predictor (e.g., a linear model or neural network) to estimate the inference latency or energy consumption of individual layers within a specific DNN model, such as MobileNet, VGG, and ResNet \cite{kang2017neurosurgeon, eshratifar2019jointdnn, liu2023adaptive, eshratifar2019bottlenet}. These predictors (treated as black boxes) are trained to establish direct mappings between the configurations of a DNN layer, including layer type, kernel size, and input feature map size, and its associated layer-wise latency \cite{liu2023adaptive}.
Then, the optimal partition solution is identified through an iterative search process, which targets the layer yielding the best \textit{predicted} performance, such as the lowest latency.
However, the partitioning performance of these explicit approaches often suffers from potential inaccuracies due to inherent uncertainties associated with these predictors. For instance, epistemic uncertainty may result from the pre-trained predictor being underfitted or the training data not being representative of the entire data space; and aleatoric uncertainty arises from noise and errors in system profiling and data collection.
Consequently, it is crucial to quantify the uncertainty associated with black-box predictions and to understand the trade-offs between the accuracy of DNN partitioning and the levels of uncertainty.

\begin{graybox}
\noindent\textbf{Insight3.} 
This observation advocates for the integration of \textbf{uncertainty awareness} into the design of the \textsc{CarbonCP} framework to assess the underlying uncertainty and ultimately improve the accuracy and decision-making quality of DNN partitioning solutions.
\end{graybox}

\textit{In summary, these three key insights drive our efforts to design and implement \textsc{CarbonCP}, a context-adaptive, carbon-aware, and uncertainty-aware inference framework aimed at advancing environmentally sustainable edge computing systems.}
\vspace{-0.15in}

\section{Conformal Prediction}
To address the challenge of uncertainty assessment (Insight 3), we propose to connect conformal prediction (CP) to the DNN partitioning process. 
CP is a statistical approach designed to provide reliable uncertainty estimates in predictions \cite{vovk2005algorithmic, shafer2008tutorial}. It offers multiple advantages over conventional approaches for measuring uncertainty.
First, CP allows for customization of the desired confidence level, providing the ability to manage the trade-off between accuracy of DNN partitioning and the level of uncertainty - a feature that is particularly crucial in high-performance AI inference frameworks (e.g., accuracy of DNN partitioning). 
Second, CP operates independently of the model type, allowing its application to any black-box model without the need to understand its internal workings. This attribute is particularly beneficial in scenarios where the underlying model is complex and its mechanics are not thoroughly understood.
Lastly, CP is data distribution-free and provides statistical guarantees along with reliable uncertainty estimates from finite samples, which reduces the efforts on measurement and data collection. 

In CP, decisions are presented as intervals that are guaranteed to contain the accurate result within a pre-defined target uncertainty level. Our goal is to obtain a prediction interval $\mathcal{C}$ that guarantees the inclusion of the true value $Y_{test}$ for new test data $X_{test}$ at a user-specified coverage rate of $1 - \alpha \in (0, 1)$, where $\alpha$ represents a nominal error level.
Assume we have a training set $\mathcal{D}_t$ and a calibration set $\mathcal{D}_c$, with $\mathcal{D}_t\cap\mathcal{D}_c=\emptyset$. $\mathcal{D}_t$ consists of a number of pairs of data $(X_j, Y_j)$, where $j = 1, \dots, |\mathcal{D}_t|$. $\mathcal{D}_c$ consists of a number of pairs of data $(X_i, Y_i)$, where $i = 1, \dots, |\mathcal{D}_c|$.  $(X_i, Y_i)$ is following any distribution $ \mathcal{P}$. Let $(X_{test}, Y_{test})$ denotes the test point, which is also sampled from the same distribution  $\mathcal{P}$. $\mathcal{D}_c \cup (X_{test}, Y_{test})$ satisfies exchangeability, which is weaker than i.i.d. (independent and identically distributed) condition.

We construct a confidence interval $\mathcal{C}$ based on the training dataset $\mathcal{D}_t$ and calibration dataset $\mathcal{D}_c$. Concretely, we train a predictor $\mathnormal{f}(x)=\hat{y}$ (any prediction function works), e.g. regression function or classification function, with the training dataset $(X_j, Y_j)\in\mathcal{D}_t$, such that ${f}(X_j)$ predicts the value of $Y_j$ that we expect to see at $X_j$. One of the key ideas behind conformal prediction is to construct \textit{nonconformity scores} symmetrically. We can calculate the nonconformity score $V_i = \mathcal{S} ((X_i, Y_i),{f})$ for any data point $(X_i, Y_i)\in \mathcal{D}_{c}$ in the calibration dataset $\mathcal{D}_c$, $i=1,..., |\mathcal{D}_c|$, where $\mathcal{S}(\cdot , \cdot)$ is the score function to indicate how the model $ {f}$ fits to the data $(X_i, Y_i)\in \mathcal{D}_c$, in other words, conforms to the calibration dataset $\mathcal{D}_c$. We can then establish an empirical distribution of nonconformity scores $\{V_1, V_2, ..., V_{|\mathcal{D}_c|}\}\cup{\infty}$: 
\begin{equation}\label{equ.emp_cal}
     \frac{1}{|\mathcal{D}_c|+1}\sum_{i=1}^{|\mathcal{D}_c|}\delta_{V_i}+\delta_{\infty}.
\end{equation}
For a new test point $X_{test}$, we can find a confidence interval $\mathcal{C}$, defined as
\begin{equation}
\mathcal{C}(X_{test}) = \{Y : \mathcal{S}((X_{test}, Y), f) \leq \hat{q}_{1-\alpha}\},
\end{equation}
where $\hat{q}_{1-\alpha}$ is the level $1-\alpha$ quantile of the empirical distribution of nonconformity scores in (\ref{equ.emp_cal}), i.e., 
\begin{equation}
    \hat{q}_{1-\alpha}=\text{Quantile} (1-\alpha,\frac{1}{|\mathcal{D}_c|+1}\sum_{i=1}^{|\mathcal{D}_c|}\delta_{V_i}+\delta_{\infty}).
\end{equation}
Then, $\mathcal{C}(X_{test})$ satisfies

\begin{equation}\label{equ.cp_guarantee}
\mathbb{P}(Y_{test} \in \mathcal{C}(X_{test})) \geq 1 - \alpha.
\end{equation}

The confidence interval $\mathcal{C}(X_{test})$ constructed by conformal prediction is guaranteed with the assumption of exchangeability  in the data. With such assumption, conformal prediction furnishes a mechanism to quantify the uncertainty of the prediction in a statistically rigorous framework. However, when test data $(X_{test}, Y_{test})$ experiences  a covariate shift from the calibration dataset $\mathcal{D}_c$ \cite{tibshirani2019conformal}, i.e.,
\begin{equation}
(X_i, Y_i) \stackrel{\text{i.i.d.}}{\sim} \mathcal{P} = \mathcal{P}_{X} \times \mathcal{P}_{Y|X}, \quad i = 1, \ldots,|\mathcal{D}_c|,
\end{equation}
\begin{equation}
(X_{test}, Y_{test}) \sim \widetilde{\mathcal{P}} = \widetilde{\mathcal{P}}_{X} \times \mathcal{P}_{Y|X}, \quad \text{$\widetilde{\mathcal{P}}_{X}$ is different from ${\mathcal{P}}_{X}$.}
\end{equation}
the generated confidence interval and its stochastic guarantee (\ref{equ.cp_guarantee}) does not hold.
In a nutshell, the test and calibration covariate distributions differ (as known as $\textit{covariate shift}$). This compromises the reliability of the confidence intervals constructed by vanilla CP. In such case, following \cite{tibshirani2019conformal}, we have to re-weight the distribution of $V_i$ in (\ref{equ.emp_cal}) to align  the calibration input data distribution $X_{test}\sim\mathcal{P}_{X}$ with the test data dsitribution  $x\sim\widetilde{\mathcal{P}}_{X}$. Specifically, we can obtain a re-weighted empirical distribution of nonconformity scores at calibration dataset \cite{tibshirani2019conformal}
\begin{equation}
\sum_{i=1}^{|\mathcal{D}_c|} p_{i}^{w}(X_{test}) \delta_{{V_i}} + p^{w}_{|\mathcal{D}_c|+1}(X_{test}) \delta_{\infty},
\end{equation}
where the weights are defined as
\begin{equation} \label{weights1}
p_{i}^{w}(X_{test}) = \frac{w(X_i)}{\sum_{i=1}^{|\mathcal{D}_c|} w(X_i) + w(X_{test})}, i = 1,\dots,|\mathcal{D}_c|,
\end{equation}
\begin{equation} \label{weights2}
p_{|\mathcal{D}_c|+1}^{w}(X_{test}) = \frac{w(X_{test})}{\sum_{i=1}^{|\mathcal{D}_c|} w(X_i) + w(X_{test})},
\end{equation}
with $w(x)=\frac{\widetilde{\mathcal{P}}_X(x)}{\mathcal{P}_X(x)}$.

 Given a nominal error level $\alpha$ and a  test point $X_{test}\sim \widetilde{\mathcal{P}}_X$, the weighted confidence bond of conformal prediction for this new test point $X_{test}$ will be 

\begin{equation}
\mathcal{C}(X_{test}) = \{Y : \mathcal{S}((X_{test}, Y), f) \leq \hat{q}_{1-\alpha}\},
\end{equation}
\textit{where $\hat{q}_{1-\alpha}$ is}
\begin{equation}
\text{Quantile}(1-\alpha; 
\sum_{i=1}^{|\mathcal{D}_c|} p_{i}^{w}(X_{test}) \delta_{{V_i}} + p^{w}_{|\mathcal{D}_c|+1}(X_{test}) \delta_{\infty}).
\end{equation}

\section{Problem Statement}
\label{sc:ps}

Each frame captured at the edge device is fed into a DNN for inference. The DNN computations of each layer can be performed at the edge device or offloaded to the server. Per-layer inference on servers has to suffer from unpredictable communication overhead between the edge and the server, as well as dynamic changes of its own computation resource. While per-layer inference on edge devices does not incur communication overhead, it fails to meet stringent real-time requirements and generates more computation due to limited resources. 

We consider a general DNN with the number of $N$ layers, such that $L = \{L_1, L_2, \dots, L_N\}$ denotes layers in the DNN. The same DNN model deployed on edge device and server. As mentioned above, we aim to find an optimal partition solution $\hat{y}$, which specifics a best guaranteed partition point to orchestrate the distribution of computation between the edge device and server. Note that the per-layer inference mechanism provides execution semantics for DNN partitioning, i.e., partitioning computation after a specific layer represents executing the DNN to that layer on the edge device, transmitting the output of that layer to the server over the wireless network, and executing the remaining layers on the server. 

Since we consider the dynamic contexts changing for the whole system including contexts in edge device, wireless network and server, such that the changing affects the best guaranteed partition point even for the same DNN architecture. For instance, while performing DNN inference on the edge device, other programs run concurrently, such as GPU consumption driven type and CPU consumption driven type, which directly affects the DNN inference computation consumption. The wireless connection of edge device often experience high variances, which also directly affect the communication overhead. The carbon intensity varies from region to region and over time in the same region, which directly affect the carbon footprint. For server side, it typically experience diurnal load patterns, such as the GPU usage, which leads to high variance in its DNN inference computation consumption. Suffering from such dynamic contexts changing in the system, there is an urgent demand for an automatic system to intelligently
search the best point to partition the DNN to optimize the carbon footprint while taking into account end-to-end latency and edge device energy consumption.

\textbf{Modeling Inference Latency.} Once the partition is made, each
frame is processed at the edge, and then sent the intermediate result from the edge to the server, and then processed at the server. Let $t_{i}^e$ and $t_{i}^s$ be the time needed to process $L_i$ on edge and server respectively. Let $B$ be network bandwidth. We define $D_t = \{d_1, d_2, \dots, d_N\}$, such that $d_{i}$ represents the output data size of $L_i$. Note that $B$ is dynamically changed and we need to adapt such changes. We define $F_e = \{t_{1}^e, t_{2}^e, \dots, t_{N}^e\}$, $F_t = \{t_{1}^t, t_{2}^t, \dots, t_{N}^t\}$, and $F_s = \{t_{1}^s, t_{2}^s, \dots, t_{N}^s\}$ to represent inference latency at the edge,  communication latency, and inference latency at the server of each layer $L_i$.

The latencies of the three stages are characterized as follows: In the edge-computing stage, the inference latency is defined as $T_e = \sum_{i=1}^{\hat{y}} t_{i}^e$. In the wireless communication stage, the transmission latency is defined as $T_t = \frac{d_{\hat{y}}}{B}$. In the server-computing stage, the inference latency is defined as $T_s = \sum_{i=\hat{y}+1}^N t_{i}^s$.

The per frame end-to-end latency can be defined as 
\begin{equation}
    T = T_{e} + T_{t} + T_{s}.
\end{equation}

\textbf{Modeling Battery Consumption of Edge Devices.} Since the dominant contexts we consider for the edge device are relative to CPU and GPU, we can break down the total energy consumption into energy consumption of CPU $E_{cpu}$, energy consumption of GPU $E_{gpu}$, and energy consumption of others $E_{others}$. We define the power consumption of CPU, power consumption of GPU, and power consumption of others as $p_{cpu}$, $p_{gpu}$, and $p_{others}$ respectively. Note that the $p_{cpu}$, $p_{gpu}$, and $p_{others}$ are dynamically changed and we need to adapt such changes.


The energy consumption for each of the three components is defined as follows: the CPU energy consumption is $E_{cpu} = p_{cpu} \cdot T_e$, the GPU energy consumption is $E_{gpu} = p_{gpu} \cdot T_e$, and the energy consumption for other components is $E_{others} = p_{others} \cdot T_e$. Hence, the per frame processing energy consumption on the edge device can be defined as 
\begin{equation}
    E_{e} = E_{cpu} + E_{gpu} + E_{others}.
\end{equation}

\textbf{Modeling Operational Carbon Footprint.} We consider the carbon footprint of the whole system, which includes the carbon footprint of the processing on the edge device, transmission, and server. Let $ci_{s}$ denote the current carbon intensity of the local area of the system. In reality, the carbon intensity can vary across different geographical locations and during different seasons/time. For carbon footprint of processing in area $s$, we can calculate the overall carbon emissions of the edge computing system as
\begin{equation}
    C = (E_e + E_t + E_s)\cdot ci_s.
\end{equation}

The variance of the system processing contexts reiterates the need for a context-adaptive solution. To better understanding DNN computation partitioning with dynamic contexts changing, we formulate the problem of searching the partitioning point as a multi-objective optimization problem. The objective is to minimize the per frame carbon footprint of a end-to-end system while satisfying the user preference of each. We introduce three positive weight parameters $\lambda_1$,  $\lambda_2$ and $\lambda_3$ to characterize the user preference of each component. We adopt the weighted sum method to formulate the optimization problem as 
\begin{equation}\label{objectivefunction}
    \mathbb{P}_0: \min_{\hat{y}} Q = \lambda_1T + \lambda_2E_e  + \lambda_3C.
\end{equation}

\section{\textsc{CarbonCP} Design}
\label{sc:carboncp}

\begin{figure}[t]
\centerline{\includegraphics[width=0.48\textwidth]{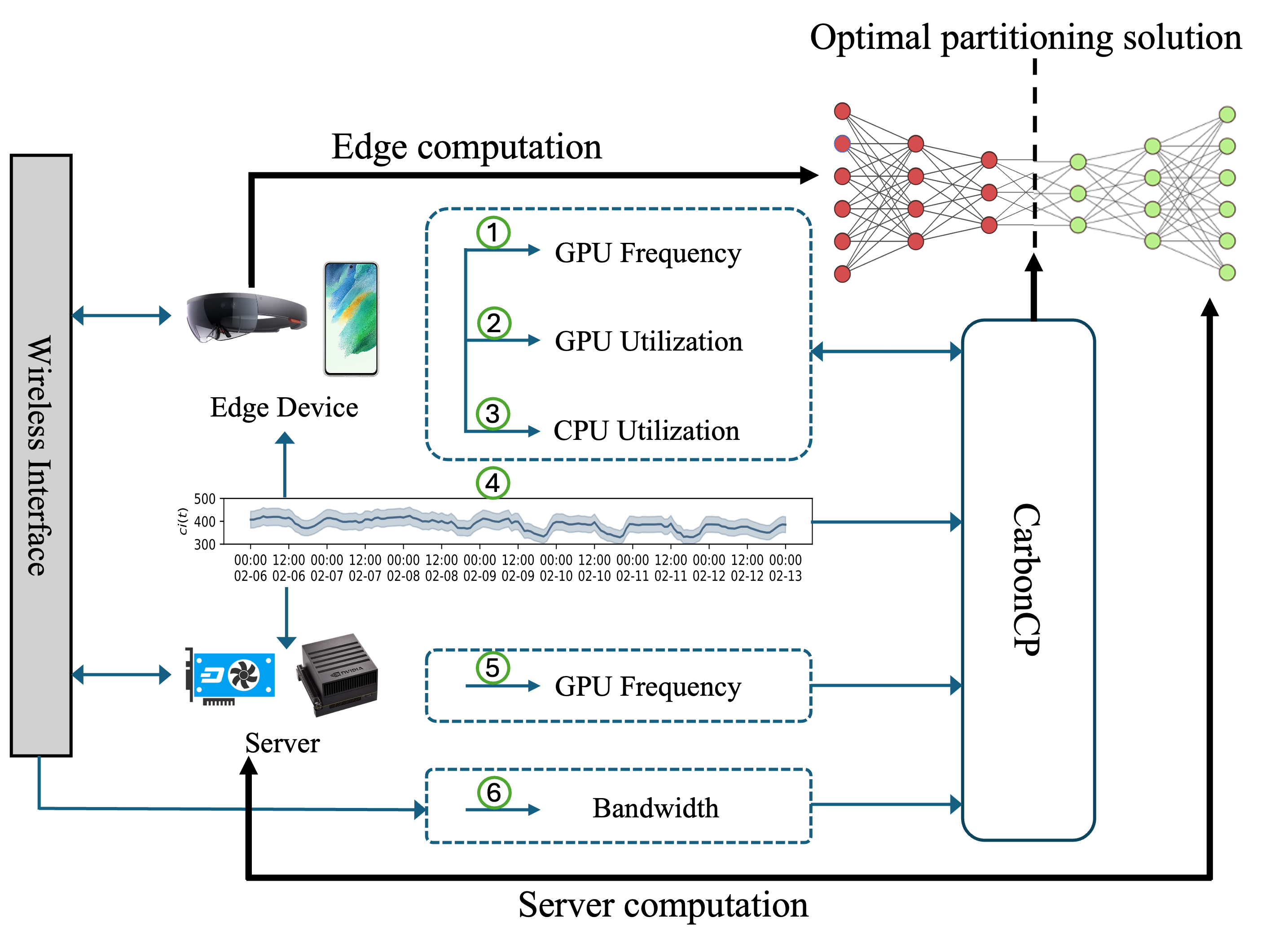}}
\caption{Overview of the proposed \textsc{CarbonCP} framework.}
\vspace{-0.15in}
\label{fig:systemOverall}
\end{figure}

As shown in the previous section, problem $\mathbb{P}_0$ is a integer non-linear programming problem. In this section, we present the proposed \textsc{CarbonCP}, a context-adaptive, carbon-aware, and uncertainty-aware inference framework based on CP, to solve the DNN partition problem formulated in Section \ref{sc:ps}.

In order to utilize the CP theory, we need to design the predictor model, score function, and confidence interval. Note that the design of these three modules is critical to the success of CP-based approaches. Our design well captures dynamic system processing contexts and the key components of the DNN partition problem without including redundant information. The core of \textsc{CarbonCP} is an interval, which runs a CP algorithm to find a set naturally encodes the model’s uncertainty about any particular input.

\newcommand\mycommfont[1]{\footnotesize\ttfamily\textcolor{blue}{#1}}
\SetCommentSty{mycommfont}

\SetKwInput{KwInput}{Input}                
\SetKwInput{KwOutput}{Output}              
\begin{algorithm}[t]
\label{algorithm:CP_covariate-shift}
\DontPrintSemicolon
  
  \KwInput{A new data $X_{test}$, a desired nominal error level $\alpha$}
  \KwOutput{Confidence interval $\mathcal{C}(X_{test})$}
  
  \KwData{Dataset $\mathcal{D}_t = \{(X_j, Y_j, Q_{j})\}, \ j = 1, \dots, |\mathcal{D}_t|$ 
  \newline
         Dataset $\mathcal{D}_c = \{(X_i, Y^{opt}_{i})\}, i = 1, \dots, |\mathcal{D}_c|$
  

  }
  
  Train a predictor model $Q_{\theta}(X, Y)$ with $\mathcal{D}_t$ to predict the corresponding $Q$,

  Take $Q_{\theta}(X, Y^{opt})$ as the score function to calculate nonconformity scores $V_i$ for each pair of calibration set $\mathcal{D}_c$

  
  
   
   
  

  Add $\infty$ to the list of $V_i$, and rank list $V_i$ in non-decreasing order,

  Estimate the distributions of calibration and test covariate points $\mathcal{P}$, $\widetilde{\mathcal{P}}$,

  Calculate the  probability proportional ratio $w(X_{test}), p_{i}^{w}(X_{test}), \text{and } p_{|\mathcal{D}_c|+1}^{w}(X_{test})$,
  
  Calculate error level $\alpha$ quantile $\hat{q}_{1-\alpha}$ of the weighted distribution $\sum_{i=1}^n p_{i}^{w}(X_{test}) \delta_{{V_i}} + p^{w}_{|\mathcal{D}_c|+1}(X_{test}) \delta_{\infty}$, such that 
$\hat{q}_{1-\alpha} = \text{Quantile}(1-\alpha; \sum_{i=1}^n p_{i}^{w}(X_{test}) \delta_{{V_i}} + p^{w}_{|\mathcal{D}_c|+1}(X_{test}) \delta_{\infty})\}$,

   $\mathcal{C}(X_{test})$ $=$ [$\hspace{0.1cm}$],
   
   \For{n in range($|DNN|$ + 1)}{
   
    \uIf{$Q_{\theta}(X_{test}, Y_n) \leq \hat{q}_{1-\alpha}$}{
      $\mathcal{C}(X_{test}).append([Q_{\theta}(X_{test}, Y_n), Y_n])$;
    }

  }

   $\textbf{Return}$ $\mathcal{C}(X_{test})$
  
\caption{\textsc{CarbonCP} Algorithm}
\end{algorithm}

\subsection{Implicit Predictor Model}

We exhaustively collected relevant dynamic contexts and the corresponding values for target variables, training dataset $\mathcal{D}_t$ and calibration dataset $\mathcal{D}_c$. Nevertheless, the prediction accuracy is highly affected by the dynamic contexts. The better the predictor $f_\theta(x)$ (from the proper training set), the tighter the prediction interval will be.

We intend to fit a predictor neural network on the training dataset $\mathcal{D}_t$, such that $f_\theta(x) = \hat{y}$, where $x$ represents the system contexts, and $\hat{y}$ is the optimal index of the DNN partitioned layer. However, we noticed that one needs to provide the label of the optimal partitions in order to train such neural network, which could be challenging because the optimal label might not be directly available. Therefore, instead of learning such explicit particition predictor model $f_\theta$,  we could adopt an implicit approach where for each set of system contexts $x$, we consider each DNN layer $y$ as a potential partitioning point ($y$ is not the optimal partitioning point). As shown in line 1 of Algorithm 1, we then fit a predictor $Q_{\theta}$ on the ($x$, $y$) and use it to predict its corresponding $Q$ which is defined in equation (\ref{objectivefunction}). The optimal partitioning point $\hat{y}$ is then determined as the value of $y$ that minimizes the predicted $Q$ value, formulated as:
\begin{equation}
    \hat{y} = \arg \min_{y} Q_{\theta}(x,y).
\end{equation}
This formulation allows us to circumvent the need for explicit labels of the optimal partitions by instead leveraging the predictions of the predictor to determine the most effective partition point based on the predicted performance metric $Q$. To address the complexity and non-linear relationships present in our high-dimensional $\mathcal{D}_t$, instead of building linear models to estimate each layers' performance, we utilize the neural network as our predictor.

\subsection{Score Function}

The score function $S(\cdot, \cdot)$ in conformal prediction is used to measure the level of non-conformity of the predicted partition decision for the calibration dataset $\mathcal{D}_c$. Moreover, unlike $\mathcal{D}_t$, our collected $\mathcal{D}_c$ includes only the optimal partitioning data $Y^{opt}$. Formally, we can calculate nonconformity score such that 
\begin{equation}
    V_i = \mathcal{S}((X_i, Y^{opt}_i), Q_\theta), \quad (X_i, Y^{opt}_i)\in\mathcal{D}_c,
\end{equation}
\noindent where $Y^{opt}_i$ is the optimal partitioning point for its corresponding system contexts $X_i$.

Intuitively, the expression $V_i$ can have the same mathematical meaning as the loss $Q_{\theta}(\cdot,\cdot)$, as in fact, a large value indicates that the model assigns a low probability to data $(X_i, Y^{opt}_i)$. So that we could directly use $Q_\theta$ to be our score function as mentioned in line 2 of Algorithm 1, such that
\begin{equation}
    \mathcal{S}((X_i, Y^{opt}_i), Q_{\theta}) =  Q_{\theta}(X_i,Y^{opt}_i).
\end{equation}
Then, we iterate the $\mathcal{D}_c$ to calculate the nonconformity score of each point. 

\begin{figure}[t]
\centerline{\includegraphics[width=0.43\textwidth]{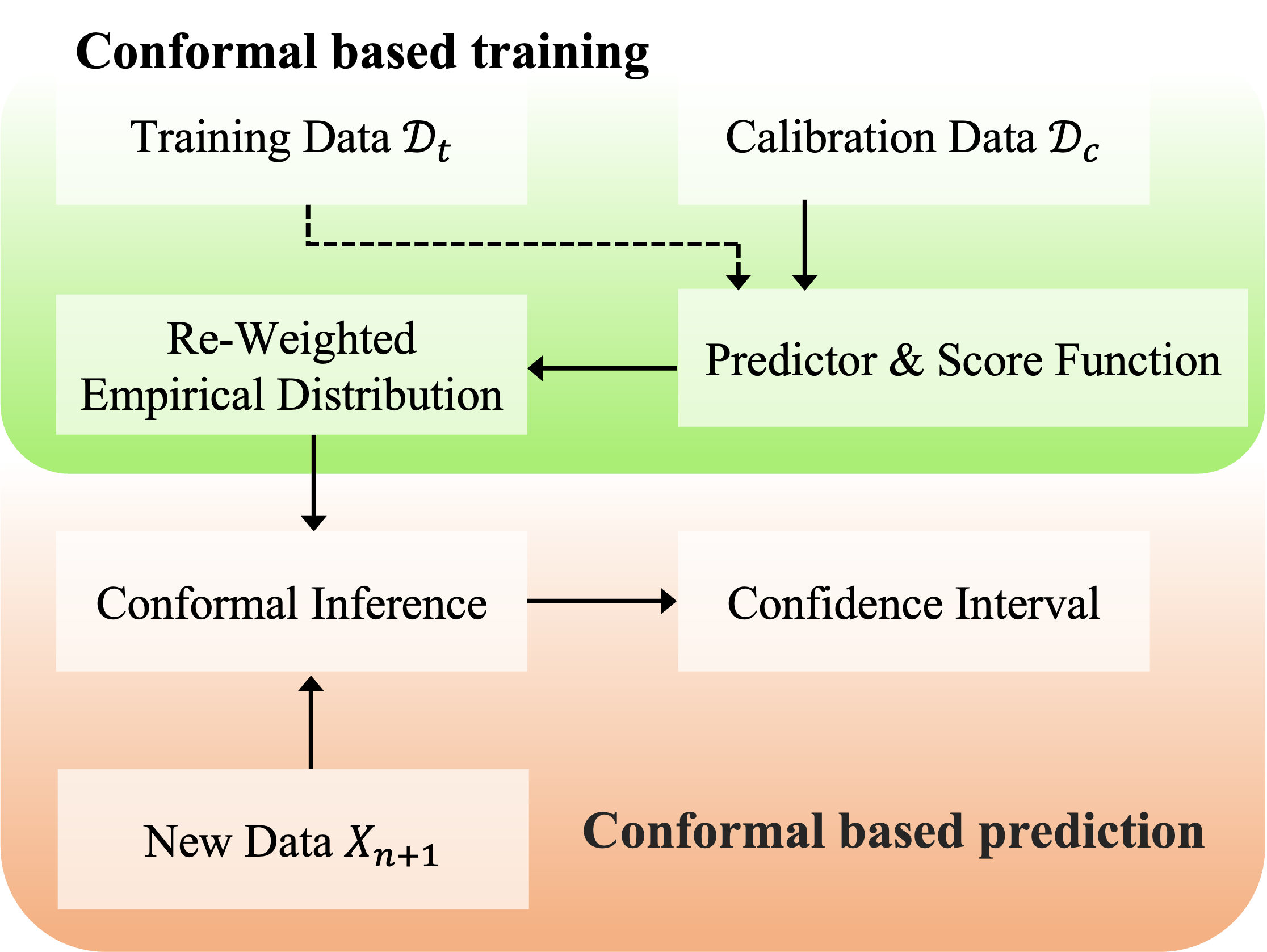}}
\caption{Simplified \textsc{CarbonCP} workflow.}
\vspace{-0.15in}
\label{fig:revise}
\end{figure}

\subsection{Confidence Interval}
\label{ssc:confidence}

Next, by sorting the nonconformity score $V_i$ in non-decreasing order, and by adding $\infty$ to the nonconformity score list (line 3 in Algorithm 1), for any new test data $X_{test}$, a matching confidence interval can be subsequently created based on the user specified nominal error level $\alpha$ such that 
\begin{equation}
    \mathcal{C}(X_{test}) = \{Y : Q_\theta(x,y) \leq  \hat{q}_{1-\alpha} \}.
\end{equation}

Even though vanilla CP has been proved to work well on establishing valid prediction intervals, our experimental results, however, show that direct application of vanilla CP to the DNN partition problem does not lead to satisfying performance (Section VI). We suspect this is due to the following two reasons: 1) There is a strong correlation between the test error of the predictor model and the average length of the prediction interval (mentioned in \cite{lei2018distribution}). 2) Our test dataset and calibration dataset are no longer exchangeable, which means  the calibration data is drawn i.i.d. from a single distribution $P_{X}$, while the test point comes from a second distribution $\widetilde{P}_{X}$. This leads to the empirical distribution of nonconformity score will not work for our case \cite{tibshirani2019conformal}. Typically, vanilla conformal prediction requires that data in $\mathcal{D}_c \cup (X_{test}, Y_{test})$ satisfies exchangeability, so that we can form a prediction interval through comparing the nonconformity score at the test point with the empirical distribution of nonconformity scores at the calibration dataset. But in practice, $\mathcal{D}_c \cup (X_{test}, Y_{test})$ are no longer exchangeable for most of cases, which is also applied in our dataset. To address these two issues, we begin by fine-tuning our prediction model including optimizing the model's hyperparameters and feature selection. Based on the work \cite{tibshirani2019conformal}, for covariate shift, we should no longer consider the empirical distribution of nonconformity scores at calibration dataset, one of the promising solution is to weight each nonconformity score by the likelihood ratio between $\widetilde{{P}}_X$ and ${P}_X$, such that $w(x) = \widetilde{{P}}_X(x) /  {{P}}_X(x)$. 
\begin{figure*}[t]
  \centering

  \begin{subfigure}[b]{0.3\textwidth}
    \centering
    \includegraphics[width=\linewidth]{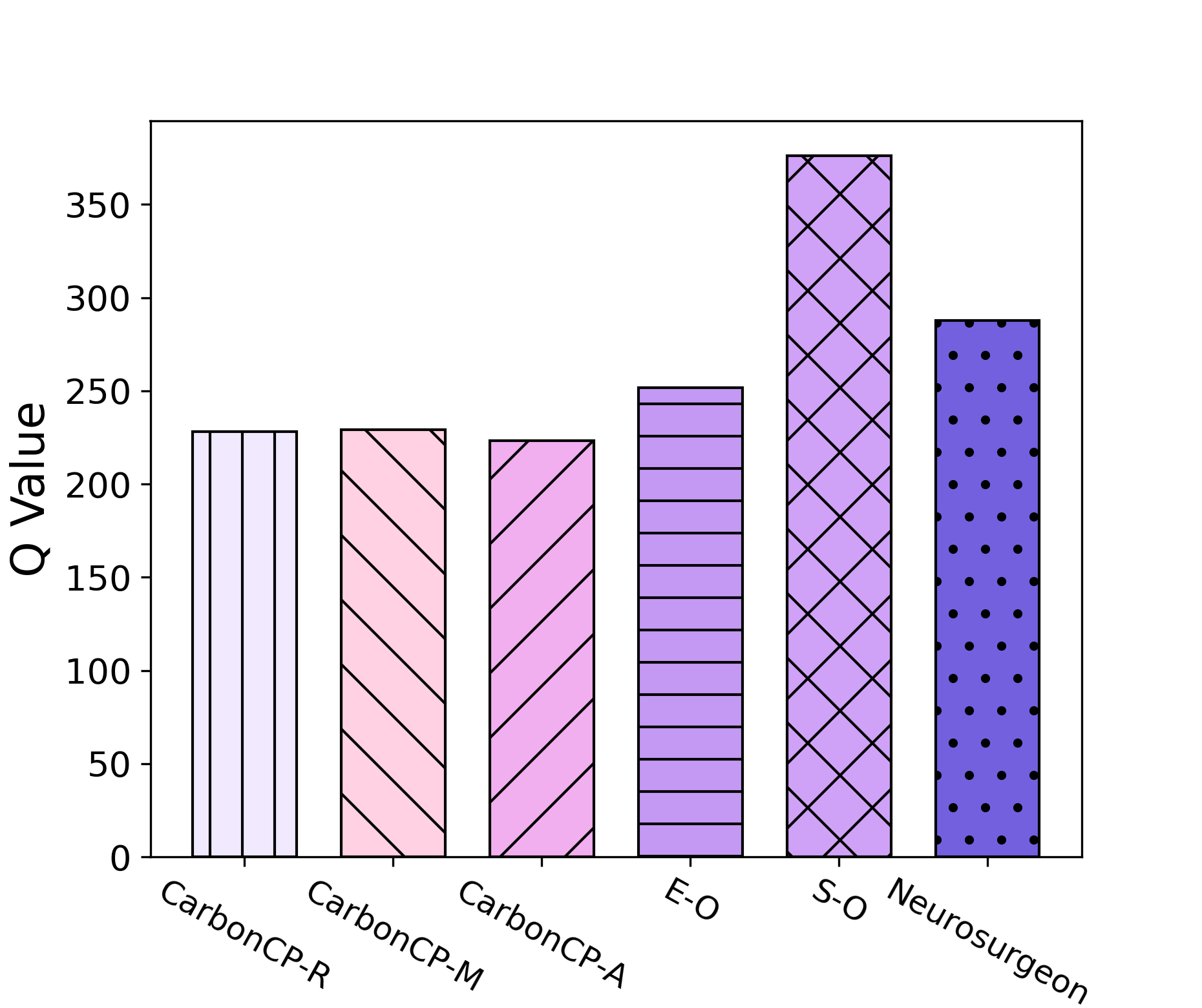}
    \label{fig:sub101}
  \end{subfigure}
  \hfill
  \begin{subfigure}[b]{0.3\textwidth}
    \centering
    \includegraphics[width=\linewidth]{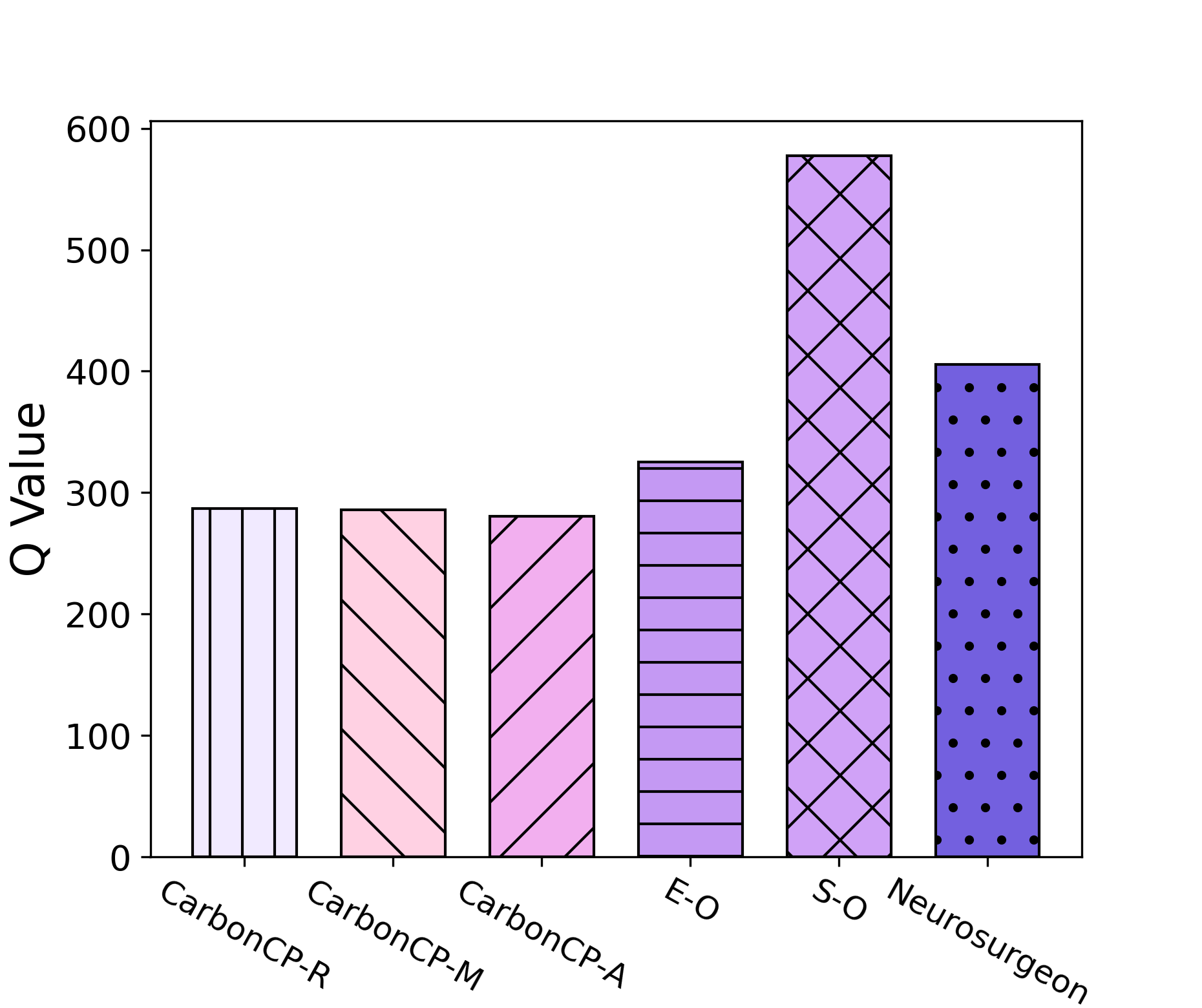}
    \label{fig:sub102}
  \end{subfigure}
  \hfill
  \begin{subfigure}[b]{0.3\textwidth}
    \centering
    \includegraphics[width=\linewidth]{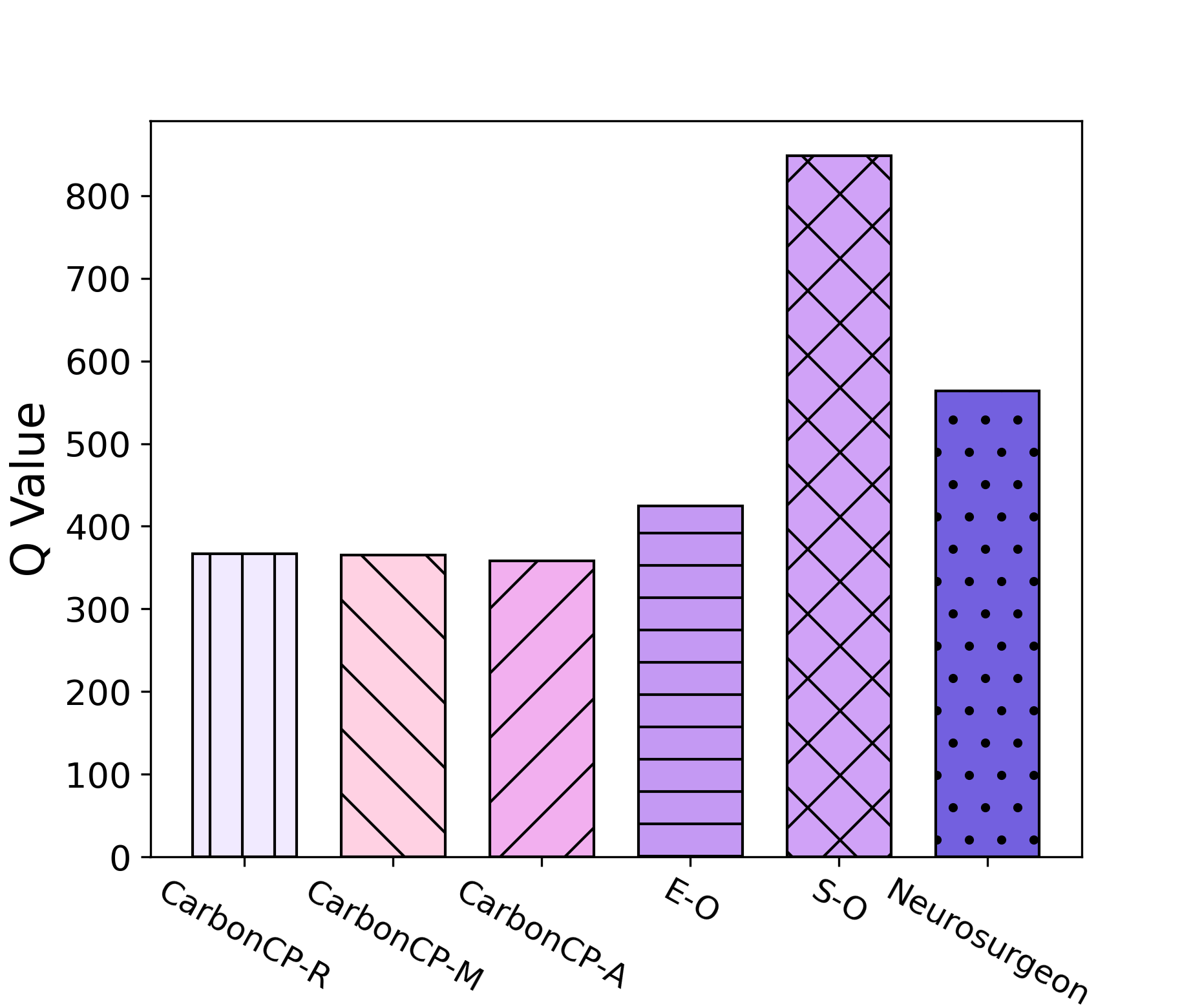}
    \label{fig:sub103}
  \end{subfigure}
  
    \vspace{-0.17in} 
    
  \begin{subfigure}[b]{0.3\textwidth}
    \centering
    \includegraphics[width=\linewidth]{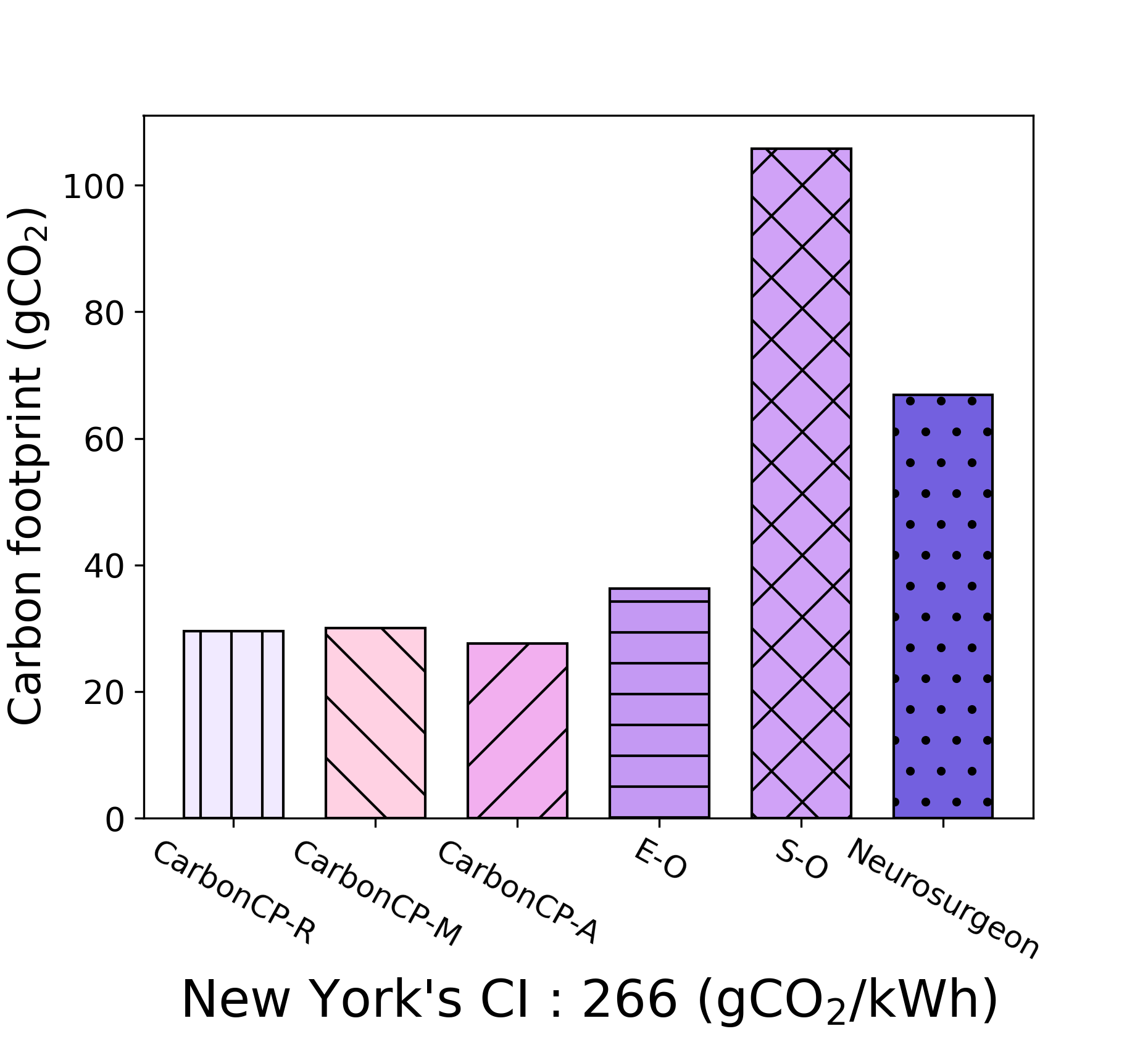}
    \label{fig:sub104} %
  \end{subfigure}
  \hfill
  \begin{subfigure}[b]{0.3\textwidth}
    \centering
    \includegraphics[width=\linewidth]{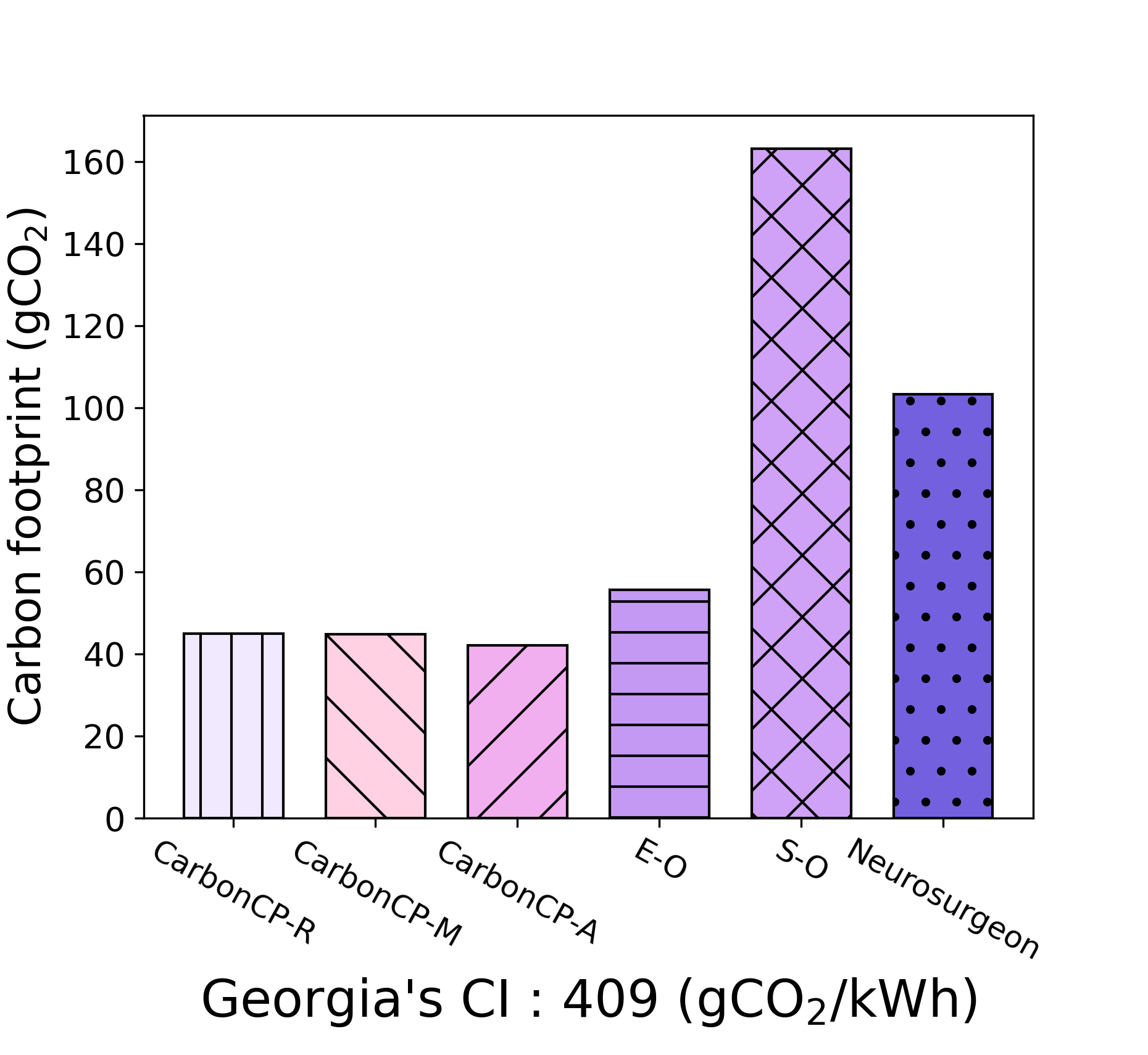}
    \label{fig:sub105} %
  \end{subfigure}
  \hfill
  \begin{subfigure}[b]{0.3\textwidth}
    \centering
    \includegraphics[width=\linewidth]{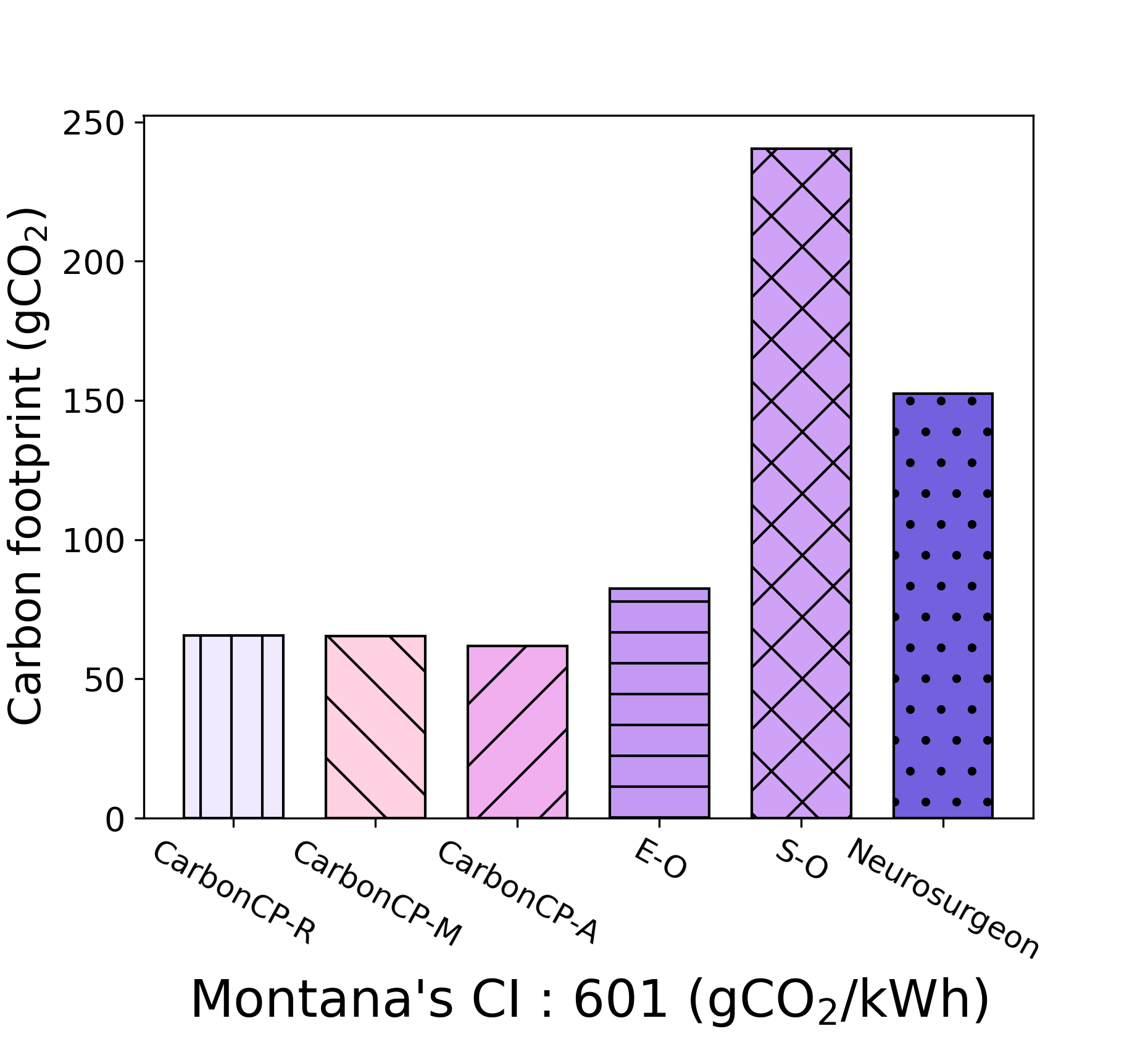}
    \label{fig:sub106} %
  \end{subfigure}
  \vspace{-0.25in}
  \caption{The impact of Carbon Intensity for different states on Q value and carboon footprint of DNN partitioning.}
  \vspace{-0.2in}
  \label{fig:Dynamic_carbon}
\end{figure*}
For exploration, we estimate the probability density function about the test data $ \widetilde{{{P}}}_X$ and calibration dataset ${P}_X$ (line 4 in Algorithm 1). The ${P}_X$ is estimated as multivariate uniform distribution since $\mathcal{D}_c$ consists of 7 context variables, each of which was exhaustively collected and is independently and uniformly distributed across its respective range.  For $ \widetilde{{{P}}}_X$, we use multivariate normal distribution to estimate it. Before system running, we do a single system contexts measurement to parameterize $ \widetilde{{{P}}}_X$.  So that we can obtain a weighted distribution of nonconformity scores at calibration dataset
\begin{equation}
\sum_{i=1}^{|\mathcal{D}_c|} p_{i}^{w}(X_{test}) \delta_{{V_i}} + p^{w}_{|\mathcal{D}_c|+1}(X_{test}) \delta_{\infty},
\end{equation}

\noindent where the weights $p_{i}^{w}(x)$ and $p_{|\mathcal{D}_c|+1}^{w}(x)$ are defined in equation (\ref{weights1}) and (\ref{weights2}).

 Given a nominal error level $\alpha$ and a new test data $X_{test}$, the weighted confidence bond of conformal prediction for this new test data $X_{test}$ will be (line7-10 in Algorithm 1)


\begin{equation}
\mathcal{C}(X_{test}) = \{Y : Q_{\theta}(X_{test},Y) \leq  \hat{q}_{1-\alpha} \},
\end{equation}

\noindent where $\hat{q}_{1-\alpha}$ is
\begin{equation}
\text{Quantile}(1-\alpha; \sum_{i=1}^{|\mathcal{D}_c|} p_{i}^{w}(X_{test}) \delta_{{V_i}} + p^{w}_{|\mathcal{D}_c|+1}(X_{test}) \delta_{\infty}).
\end{equation}

The overall \textsc{CarbonCP} framework and workflow are described in Algorithm 1 and Fig. \ref{fig:revise}. We first fit a implicit predictor $Q_{\theta}$ on $\mathcal{D}_t$ to predict $Q$ for each DNN layer $y$ under the system contexts $x$. Then we use this $Q_{\theta}$ as the score function to calculate nonconformity scores of each optimal data in $\mathcal{D}_c$. After that, we weight each nonconformity score and get the desired $\hat{q}_{1-\alpha}$. Lastly, for any new test data $X_{n+1}$, its confidence interval can be get through line 7-10 in Algorithm 1.


\subsection{Assess Prediction Interval Performance}

When working with the confidence interval from conformal prediction, which contain multiple possible true values, the challenge is choosing a single best prediction. To transform the uncertainty of the confidence interval into a concrete basis for decision-making, we utilize this interval through three approaches. The first approach is to take a random point of the interval as the single predictive value, represented as \textsc{CarbonCP}-R. The second approach to take the mean of the interval as the single predictive value, represented as \textsc{CarbonCP}-M. The third approach is to adaptively take the point from the interval that minimizes the objective function value, represented as \textsc{CarbonCP}-A. 

Fig. \ref{fig:systemOverall} shows the overview of our proposed system. In the first step, the edge device send the service requests to build the connection with server.  To determine the bandwidth between an edge device and a server, we utilize the tool “ping” at edge device,  transmitting two data packets of different sizes to the server in succession, and measure the response times. The bandwidth equals to the
ratio between the difference of data size and the difference of response times. In the second step, the server continuously sends its GPU utilization to edge device. In the third step, after the edge device combines itself current contexts with server's contexts, carbon intensity and bandwidth, taking those system contexts as the input for \textsc{CarbonCP} to obtain the partitioning point. Then the edge device sends the partitioning point and intermediate results to the server. In the final step, the server returns the DNN inference results to the edge device.

\section{Performance Evaluation}

In this section, we evaluate both the proposed carbon-aware DNN partition model and CP algorithm. 
We implement a context-adaptive prototype system to execute Resnet-18 model. We use the NVIDIA Jetson Nano Developer Kit as the edge device, integrated with a Linksys WUSB6300 USB WiFi adapter. We use the server with a single GPU GeForce RTX 3080Ti 12GB. We employ WiFi as the communication link between edge device and the server. We use Linksys WRT1900AC as the WiFi rounter. We implement our client-server interface utilizing gRPC, an open source flexible RPC interface for inter-process communication. For flexibility in dynamically selecting partition points, both edge devices and server host complete DNN models.

The duty of edge device is to 1) receive the current system contexts of the server and estimate the real-time network bandwidth, 2) make partition decision, 3) execute the layers allocated to the edge device, 4) send partition decision and the intermediate result to server.
The duty of server is to 1) receive the partition decision and the intermediate result from edge device, 2) execute the layers allocated to the server.

\begin{figure}[t]
  \centering

  \begin{subfigure}[b]{0.48\textwidth}
    \centering
    \includegraphics[width=\linewidth]{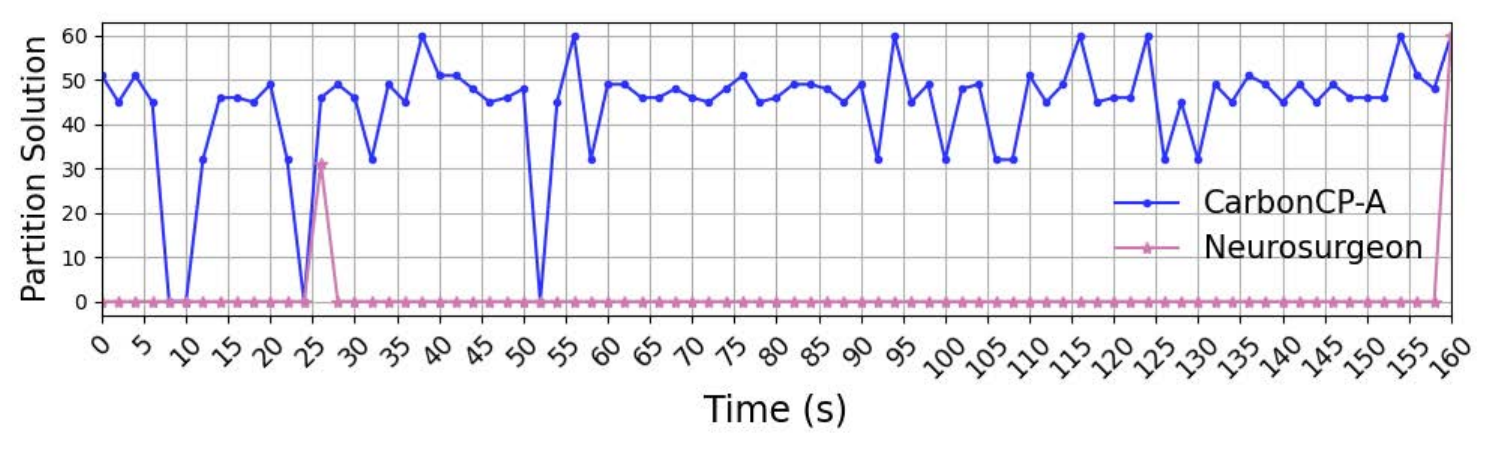}
    \caption{Partition solutions over time.}
    \label{fig:sub15}
  \end{subfigure}
  \hfill
  \begin{subfigure}[b]{0.48\textwidth}
    \centering
    \includegraphics[width=\linewidth]{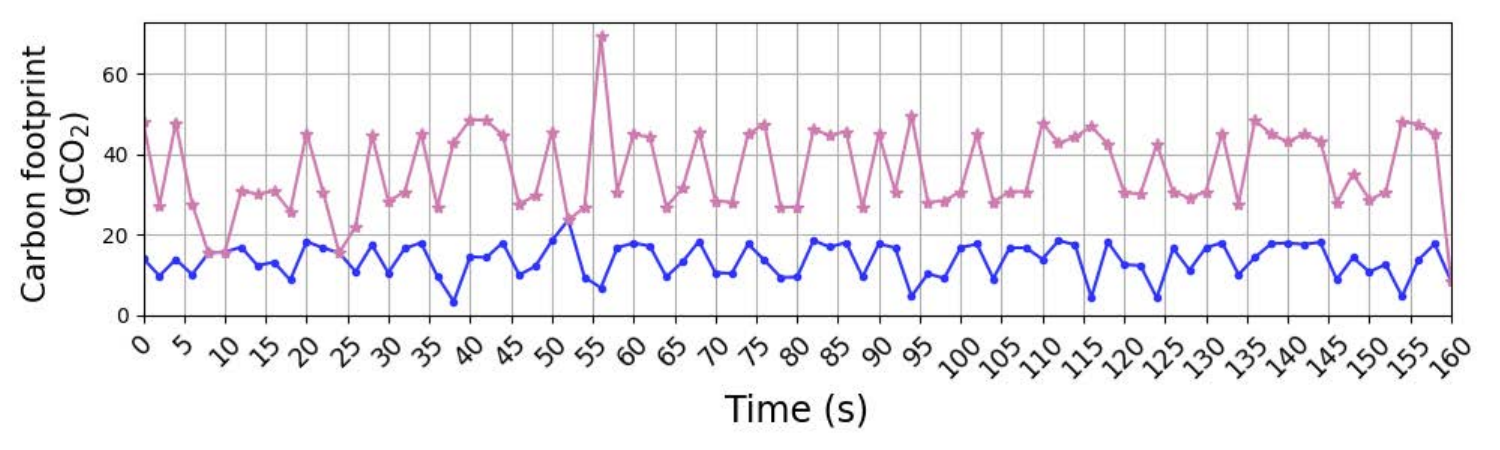}
    \caption{Operational carbon emissions over time.}
    \label{fig:sub16}
  \end{subfigure}
   \hfill
  \begin{subfigure}[b]{0.48\textwidth}
    \centering
    \includegraphics[width=\linewidth]{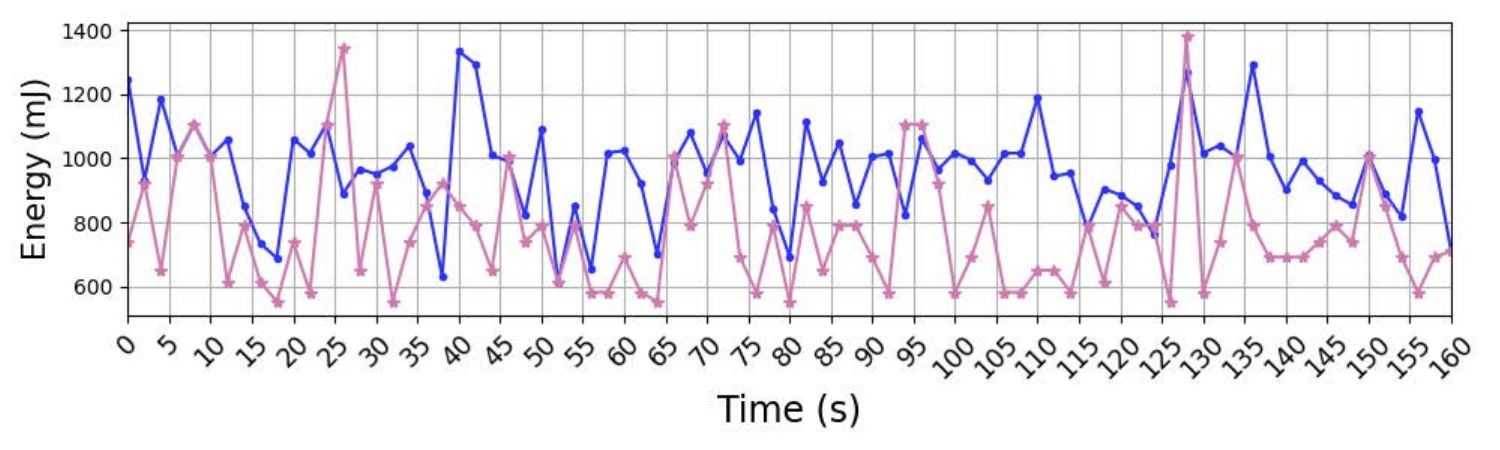}
    \caption{Battery consumption over time.}
    \label{fig:sub17}
  \end{subfigure}
  \caption{The Comparison between \textsc{CarbonCP}-A and Neurosurgeon for different metrics in real system.}
  \vspace{-0.15in}
  \label{fig:performance}
\end{figure}

\vspace{-0.15in}
\subsection{Dataset Construction}

To evaluate \textsc{CarbonCP}, we need the training dataset $\mathcal{D}_t$ and the calibration dataset $\mathcal{D}_c$ that can concurrently provide the dynamic contexts data when inferencing, including bandwidth, server GPU utilization, client GPU utilization, client GPU frequency, client CPU utilization, partition point, carbon intensity, end-to-end latency, client energy, communication energy, and server energy. To the best of our knowledge, we are not aware of any existing dataset that satisfies all these requirements. To this end, we conduct a measurement study and generate system contexts datasets with sufficient contexts data and ground truth labels. Overall, we collect fine-grained and accurate data with around 7.7 million unique configurations.

\subsection{Performance Comparison}

We compared our CP-based context-adaptive framework with two widely used baseline solutions as well as Neurosurgeon: 

\begin{itemize}
\item \textbf{Carbon-R:} the partition result is randomized from the point corresponding the confidence interval.
\item \textbf{Carbon-M:} the partition result come from the point corresponding to the value nearest to the mean of the confidence interval.
\item \textbf{Carbon-A:} the partition result come from the point corresponding to the minimum value of objective function within the confidence interval.
\item \textbf{Edge-only:} the edge device performs all DNN layers locally with limited computation recourse.
\item \textbf{Cloud-only:} the server performs all DNN layers with communication.
\item \textbf{Neurosurgeon:} it can partition DNN between the edge device and server at granularity of neural network layers by directly using regression model to predict per-layer performance explicitly. But it only considers the bandwidth and server workload as dynamic changing and it targets for optimize latency and edge side energy.

\end{itemize}

We first compare our \textsc{CarbonCP} between Carbon-R, Carbon-M, and Carbon-A, and compare \textsc{CarbonCP} with Edge-only, Server-only and Neurosurgeon under the test dataset with dynamic contexts changing in Fig. \ref{fig:Dynamic_carbon}. We see that \textsc{CarbonCP} achieves a lower Q value and carbon footprint compared with other methods.

More specifically, to show \textsc{CarbonCP} can work for different area, we randomly select the carbon intensity of three states (New York, Georgia, and Montana) in our test. \textsc{CarbonCP}-A has a Q value reduction of 2.3\%, and 2.7\% compared with \textsc{CarbonCP}-R and \textsc{CarbonCP}-M respectively. \textsc{CarbonCP}-A has a Q value reduction of 13.8\%, 49.9\%, and 30.1\% compared with Edge-only, Server-only and Neurosurgeon respectively under dynamic contexts changing. \textsc{CarbonCP}-A has a carbon footprint reduction of 6.4\%, and 6.9\% compared with \textsc{CarbonCP}-R and \textsc{CarbonCP}-M respectively. \textsc{CarbonCP}-A has a carbon footprint reduction of 24.5\%, 74.1\%, and 59.2\% compared with Edge-only, Server-only and Neurosurgeon respectively under dynamic context changing shown in graph of Fig. \ref{fig:Dynamic_carbon}. This is because, Edge-only method executes the entire DNN on the edge side, it avoids data transmission and benefits the lower power consumption of edge devices. Server-only method ignores the effect of computation limitation. Neurosurgeon ignores the dynamic contexts changing of inter edge devices and server. \textsc{CarbonCP} consider both dynamic contexts of whole system and transmission, and it makes a proper trade-off between them. For the inter-comparison between \textsc{CarbonCP}, since Carbon-A iterate over each element within the confidence interval to obtain the minimum value on objective function, it is logical that Carbon-A can achieve the best possible result than others. But at the corresponding cost is the computational complexity is $O(|\mathcal{C}_n(X_{n+1})|)$, which is highly depends on the length of confidence interval. In a nutshell, these results confirms that \textsc{CarbonCP} significant outperforms Edge-only, Server-only and Neurosurgeon methods, and Carbon-A outperforms Carbon-R and Carbon-M. 

\begin{figure*}[t]
  \begin{subfigure}{0.163\textwidth}
    \includegraphics[width=\linewidth]{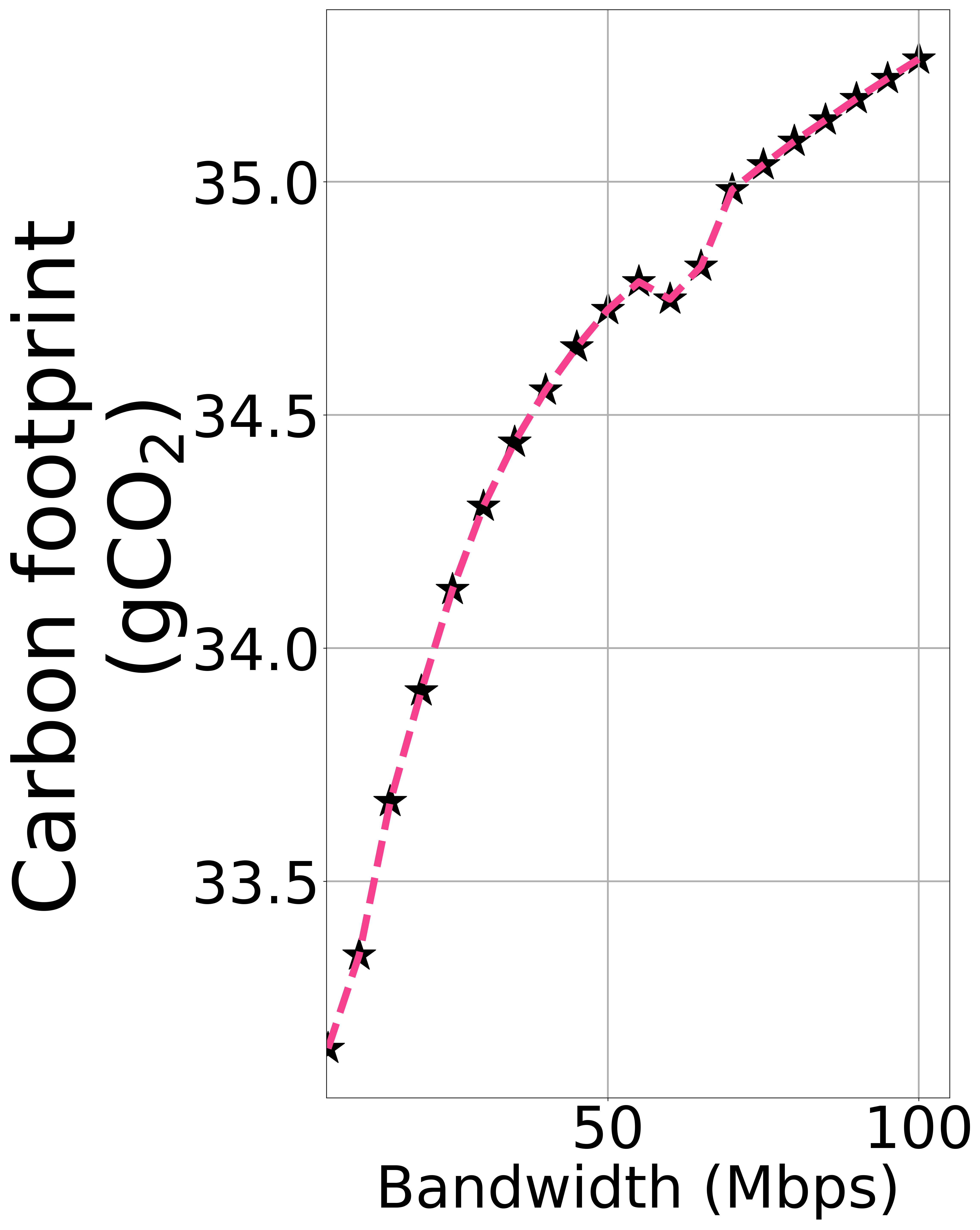}
    \label{fig:sub301}
  \end{subfigure}
  \begin{subfigure}{0.157\textwidth}
    \includegraphics[width=\linewidth]{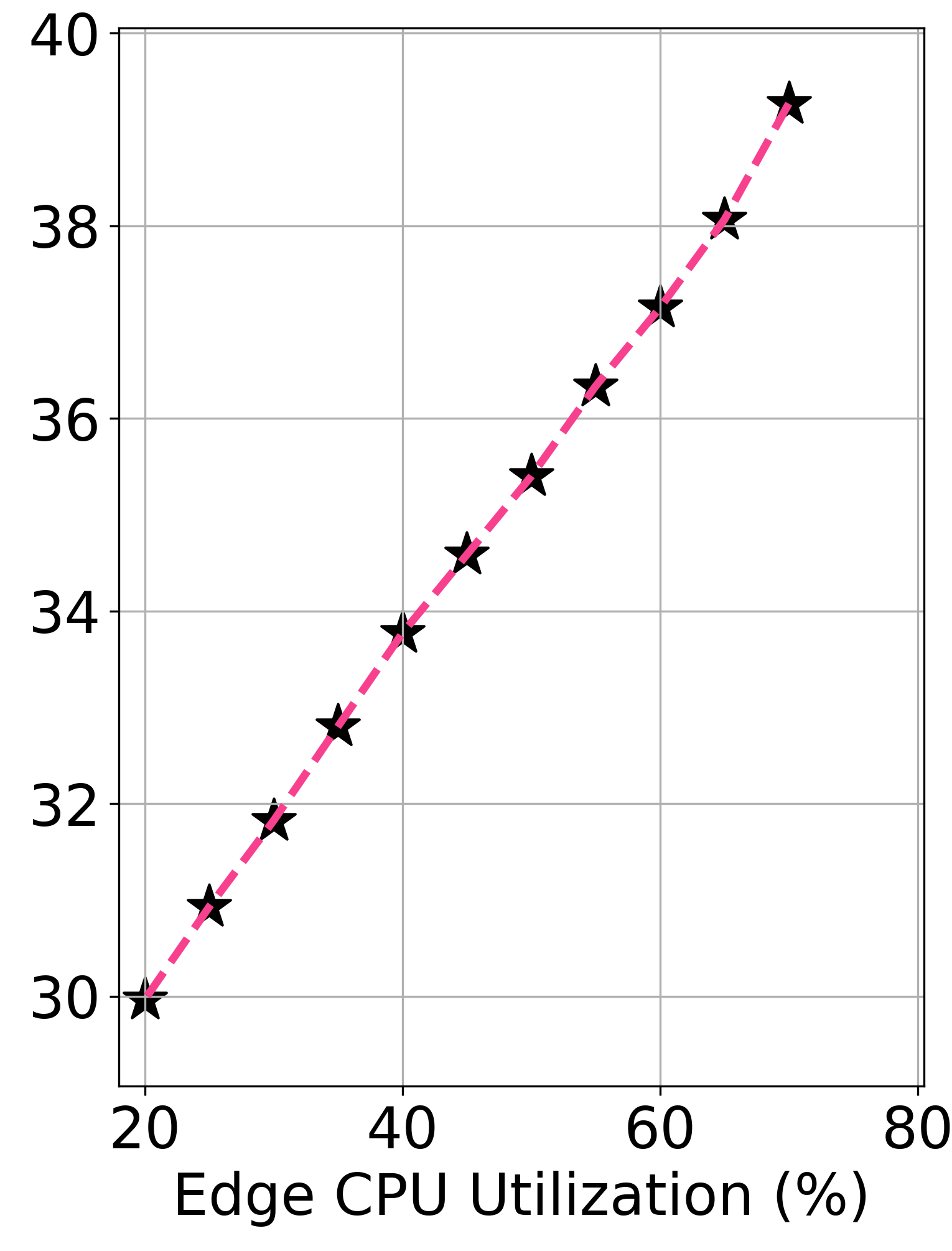}
    \label{fig:sub302}
  \end{subfigure}
  \begin{subfigure}{0.156\textwidth}
    \includegraphics[width=\linewidth]{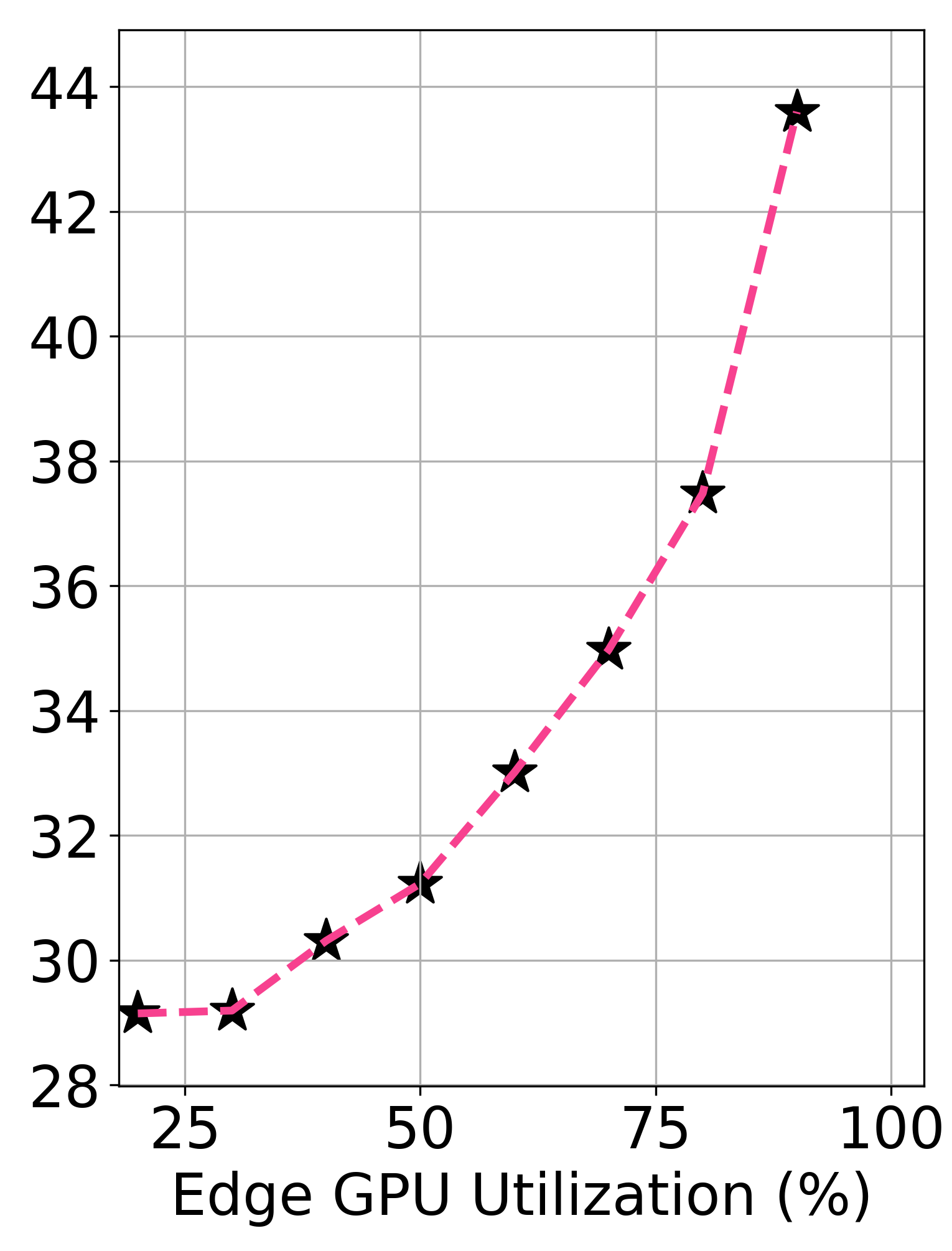}
    \label{fig:sub303}
  \end{subfigure} 
    \begin{subfigure}{0.157\textwidth}
    \includegraphics[width=\linewidth]{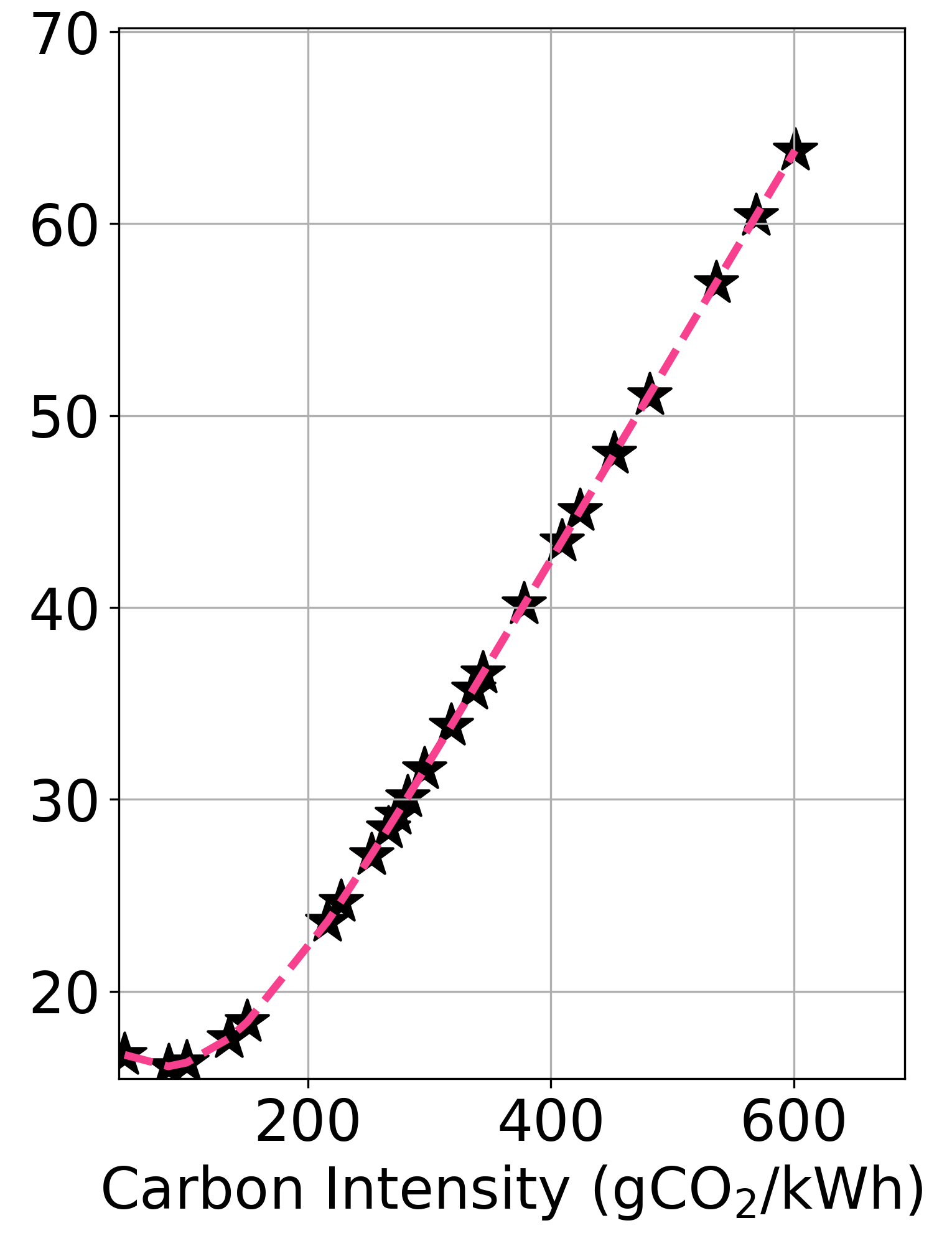}
    \label{fig:sub306}
  \end{subfigure}
  \begin{subfigure}{0.156\textwidth}
    \includegraphics[width=\linewidth]{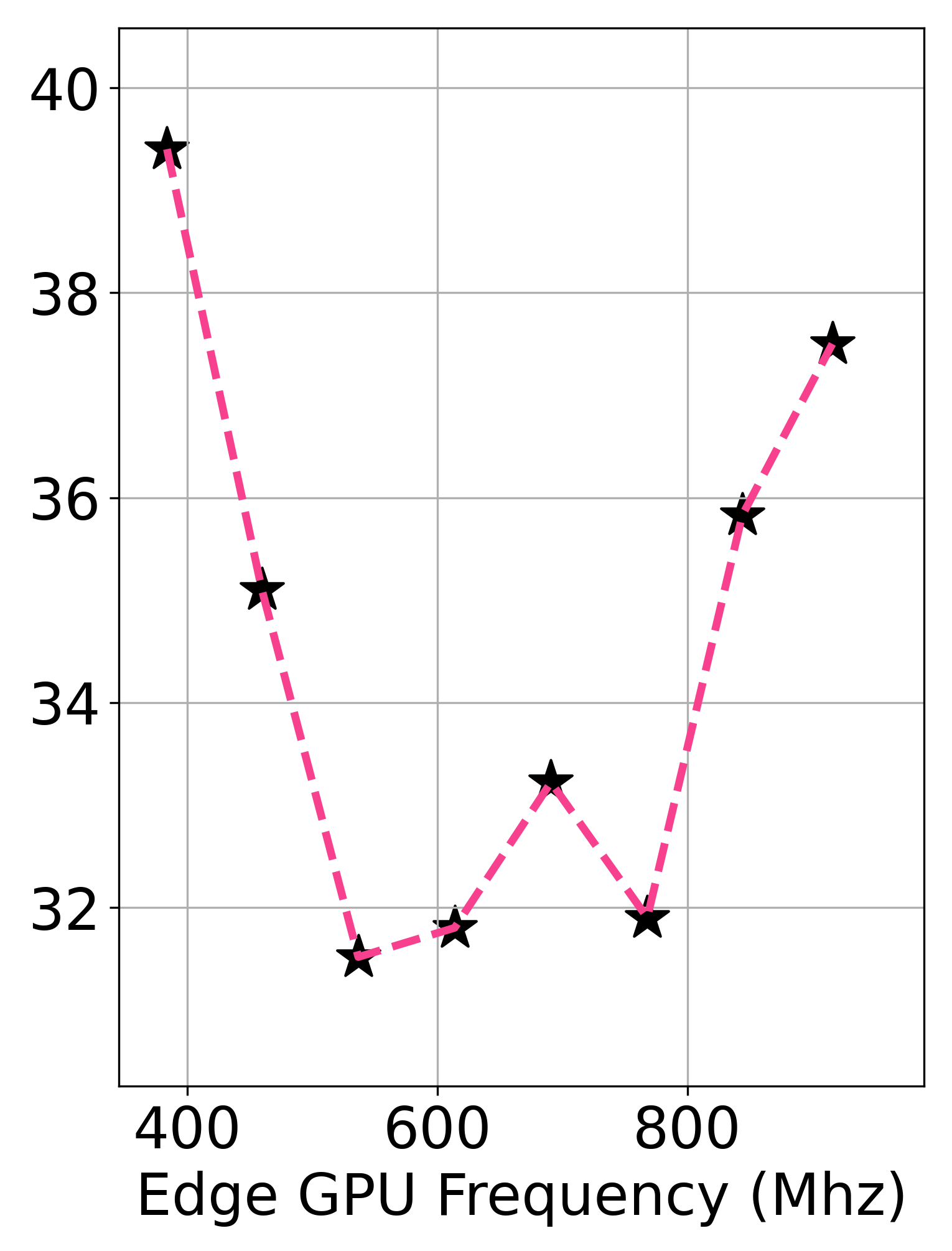}
    \label{fig:sub304}
  \end{subfigure} 
  \begin{subfigure}{0.158\textwidth}
    \includegraphics[width=\linewidth]{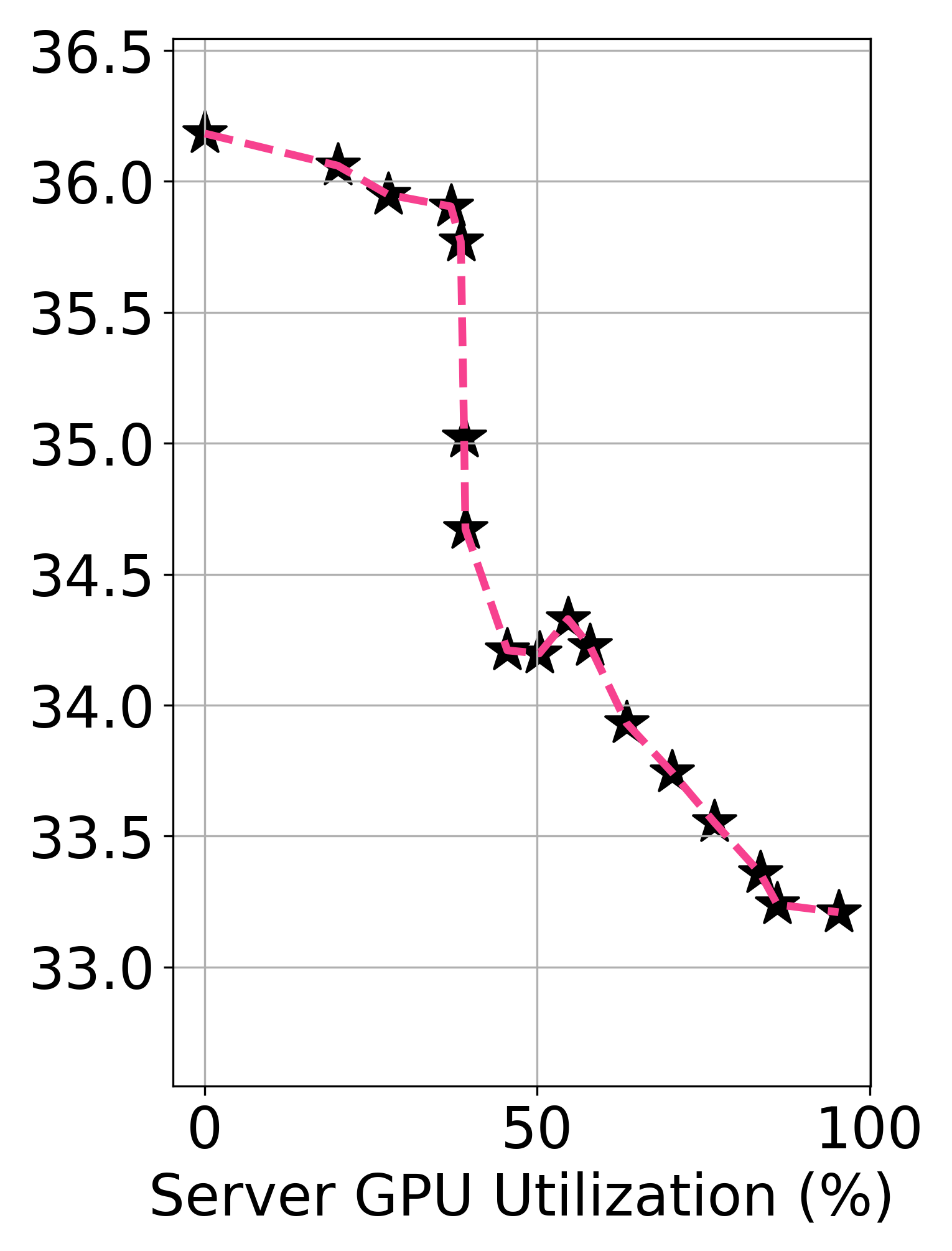}
    \label{fig:sub305}
  \end{subfigure}
  \vspace{-0.15in}
  \caption{The impact of system processing contexts and carbon intensity on overall carbon footprint.}
  \vspace{-0.2in}
  \label{fig:Single_variable}
\end{figure*}

\subsection{Performance Evaluation of \textsc{CarbonCP}-A}

We examine the performance of \textsc{CarbonCP}-A and Neurosurgeon under dynamic contexts changing in real system, as illustrated in Fig. \ref{fig:sub15}. We observe that the partition solution of \textsc{CarbonCP}-A changes dynamically and with high frequency along with time, whereas Neurosurgeon also changes the partitioning solution to some extent, but with very low frequency. This indicates that Carbon-A is context sensitive and able to adjust the partition solution as the context changes. 

In addition, we further examine the carbon footprint and energy emissions corresponding to the performance results as shown in Fig. \ref{fig:sub16} and \ref{fig:sub17} respectively. For the carbon footprint, in average, the \textsc{CarbonCP}-A has a reduction of 58.8\% compared with Neurosurgeon. We also can see, for energy emissions, the average Neurosurgeon's is 17.6\% lower than \textsc{CarbonCP}-A, which illustrates \textsc{CarbonCP}-A is relatively comparative with Neurosurgeon in energy emission. This observation validates the usefulness of \textsc{CarbonCP}-A for not only works for carbon footprint, but also consider the impact of energy emission. 

\vspace{-0.15in}
\subsection{Processing Context Variations}

In evaluating the performance of \textsc{CarbonCP}-A, we consider key metrics that include bandwidth, edge CPU utilization, edge GPU utilization, edge GPU frequency, server GPU utilization, and carbon intensity. The analysis of these metrics provides insights into their respective impacts on the carbon footprint.

For bandwidth, edge CPU utilization, edge GPU utilization, and carbon intensity as shown in Fig. \ref{fig:Single_variable}, a consistent trend emerges. These figures collectively demonstrate a direct proportionality between each metric and the carbon footprint. Specifically, augmentations in bandwidth, edge CPU and GPU utilization are accompanied by a commensurate increase in carbon emissions, indicating a robust linkage to energy consumption. Notably, carbon intensity is revealed as a paramount factor, with a stark positive correlation to carbon emissions, suggesting that the choice of carbon intensity, is vital for managing the carbon footprint.

Conversely, edge GPU frequency and server GPU utilization depict more intricate relationships in Fig. \ref{fig:Single_variable}. The edge GPU frequency presents a non-linear association with carbon footprint, intimating an optimal frequency range that minimizes emissions. This suggests that optimizing the operating frequency of GPUs will be beneficial in reducing the carbon footprint. The carbon footprint begins to decrease after the server GPU utilization increases to a certain point, potentially indicating a regime of enhanced energy efficiency at specific utilization thresholds.

Among all the analyzed factors, edge GPU utilization and carbon intensity appear to have the greatest impact on the carbon footprint. This implies that reducing GPU utilization and opting for electricity with lower carbon intensity would be the most effective measures in reducing the overall carbon footprint of \textsc{CarbonCP}-A. Moreover, the optimization of operational frequencies for GPUs emerges as an actionable measure for carbon footprint reduction. 

\vspace{-0.15in}
\subsection{Confidence Interval Utilization}
\label{ssc:interval}
\begin{figure}[t]
\centerline{\includegraphics[width=0.36\textwidth]{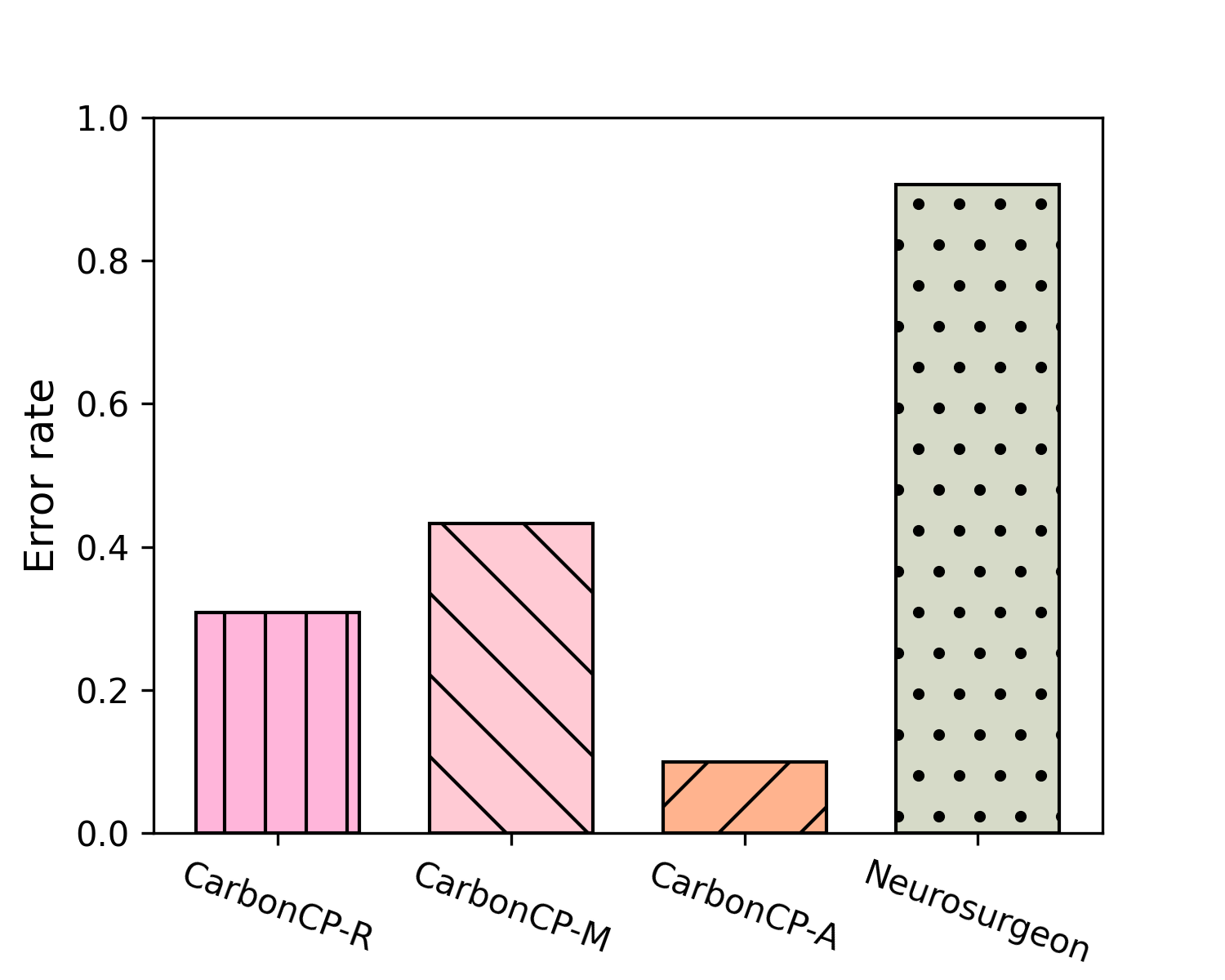}}
\caption{The error rate of \textsc{CarbonCP} and Neurosurgeon.}
\vspace{-0.3in}
\label{fig:errorRate}
\end{figure}

In this experiment, we implement \textsc{CarbonCP}-R, \textsc{CarbonCP}-M, and \textsc{CarbonCP}-A to further explore the potential utilization of the confidence interval obtained from conformal prediction. As shown in Fig. \ref{fig:errorRate}, the error rate of each CP method when applied to the test data, with the neurosurgeon's error rate included for comparison. \textsc{CarbonCP}-R and \textsc{CarbonCP}-M have relatively similar error rates 30.86\% and 43.2\% respectively, indicating a comparable level of performance. Neurosurgeon has highest error rate, 90.5\%. \textsc{CarbonCP}-A, however, shows a significantly lower error rate 9.9\%, suggesting a more accurate or calibrated method under the conditions of this test.


\section{RELATED WORK}

\textbf{Deep Neural Network partition.}
Previous research efforts focusing on offloading the resource-constrained mobile device to the powerful server/cloud will reduce inference time. Neurosurgeon \cite{kang2017neurosurgeon}, Edgent \cite{li2018edge}, and JALAD \cite{li2018jalad} are typical partition methods that split a DNN into two parts at the granularity of layers. However, those methods are not applicable for the computation partition performed by \textsc{CarbonCP} for a number of reasons: 1.) they do not consider the dynamic contexts changing of the whole system while running. 2.) they do not consider the optimization performance for the whole system carbon footprint while balancing the latency and mobile device energy. 

\noindent\textbf{Conformal Prediction.} 
To build more trustworthy machine learning models, it is significant to reliably characterize the uncertainty in their predictions \cite{sullivan2015introduction, soize2017uncertainty}. There exists a plethora of uncertainty quantification approaches such as Bayesian methods \cite{blundell2015weight, neal2012bayesian}, Monte Carlo dropout \cite{gal2016dropout}, ensembles \cite{ashukha2020pitfalls, lakshminarayanan2017simple}. However, many such methods can be limited in practice due to their computational overhead, incorrect distributional assumptions, or predisposition to specific architectures and training procedures. For example, Bayesian methods are computationally intensive and require algorithmic modifications. Ensembles require significant computational resources due to the need to train multiple models. 

\section{Conclusion}
This paper has presented a novel framework, \textsc{CarbonCP}, to improve the environmental sustainability of deploying intelligence at the network edge. \textsc{CarbonCP} offers an uncertainty-aware solution to balance the trade-offs between carbon emissions, latency, and battery consumption under dynamic processing contexts and varying carbon intensity. We hope that \textsc{CarbonCP} will instigate a fundamental reevaluation of edge computing systems, fostering a transition towards environmental sustainability.

\bibliographystyle{unsrt}
\bibliography{ref}

\end{document}